# Bandwidth Analysis of Multiport Radio-Frequency Systems—Part I

Ding Nie, *Member, IEEE,* and Bertrand M. Hochwald, *Fellow, IEEE*

*Abstract*—When multiple radio-frequency sources are connected to multiple loads through a passive multiport matching network, perfect power transfer to the loads across all frequencies is generally impossible. In this two-part paper, we provide analyses of bandwidth over which power transfer is possible. Our principal tools include broadband multiport matching upper bounds, presented herein, on the integral over all frequency of the logarithm of a suitably defined power loss ratio. In general, the larger the integral, the larger the bandwidth over which power transfer can be accomplished. We apply these bounds in several ways: We show how the number of sources and loads, and the coupling between loads, affect achievable bandwidth. We analyze the bandwidth of networks constrained to have certain architectures. We characterize systems whose bandwidths scale as the ratio between the numbers of loads and sources.

The first part of the paper presents the bounds and uses them to analyze loads whose frequency responses can be represented by analytical circuit models. The second part analyzes the bandwidth of realistic loads whose frequency responses are available numerically. We provide applications to wireless transmitters where the loads are antennas being driven by amplifiers. The derivations of the bounds are also included.

## I. INTRODUCTION

Analyses of the bandwidth over which power can be transferred in radio-frequency (RF) systems are often limited to a single source and load because of the complexity in manipulating multiport matching networks and multiple loads. Factors that contribute to the complexity include defining appropriate measures of bandwidth when there are many sources and loads, and the difficulty of analyzing coupling between loads. We propose methods of analysis that utilize broadband performance bounds applicable to a wide class of passive networks and an arbitrary number of sources and dissipative loads.

The ability to transfer power from sources to loads relies, in part, on the ability to match the impedance of the sources to the frequency-dependent impedance $Z_L(j\omega)$ of the loads over a broad frequency range. Bandwidth upper bounds are of great help in determining the best achievable bandwidth performance for a given load. Classical Bode-Fano results [1], [2] on the integral of the logarithm of the reflection coefficient can be used for such bounds when there is a single source and load. When there are multiple loads, analyses are often limited to special cases. For example, bandwidth bounds for loads with some specific structures are discussed in [3]–[6], and examples of analyses of multiport systems include [7], [8]. Often, multiple reflection coefficients are defined and analyzed separately using scalar Bode-Fano theory. However, as shown in [9], a physically-meaningful single reflection coefficient can be defined and analyzed when there are $N$ arbitrarily coupled loads driven by $N$ sources ($N > 1$).

In this two-part paper, we present a bandwidth analysis of matched multiport RF systems that builds on bandwidth bounds in [9]. The first part presents the bounds and applies them to systems that can be expressed in closed form. The second part provides proofs of the bounds and applies them to systems whose scattering parameters are expressed numerically. The bounds in [9] apply to loads that are modeled as perfect reflectors as $\omega \to \infty$. We extend these results, and present bounds that apply to loads that are reflectors at any frequency, including $\omega = 0$. We allow the matching network and loads to be non-reciprocal. The network can also be lossy. We permit the number of sources and loads to be unequal.

By applying bandwidth bounds, we demonstrate how the number of sources, loads, and the coupling between loads affect the achievable bandwidth of a matched multiport system. We prove that bandwidth bounds generally scale as $N/M$, where $M$ is the number of sources and $N$ is the number of loads. This result also holds in the presence of coupling, as long as it is not "too strong". This suggests that unlimited bandwidth is theoretically achievable by simply adding more loads for a fixed number of sources. As is shown, both the loads and the network architecture play an important role in achieving linear-in-$N$ performance of the overall system, for a given $M$.

We also propose a bandwidth analysis for situations where a portion of the network is constrained to have a certain structure while other portions are unconstrained. This situation occurs in beamforming applications since a beamforming antenna array can be thought of as $N$ loads driven with prescribed amplitude and phase relationships by a single source.

The basic premise of broadband matching is that when a source and load are connected to each other, even if the reflection coefficient is made small at a design frequency $\omega = \omega_d$, it is generally not small for all $\omega$ [10]. When there are multiple sources and loads, there is no single reflection coefficient since power sent from source $i$ may, through coupling, return to source $j \neq i$. In [9], a definition of a multiport reflection coefficient that takes this phenomenon

D. Nie is with Apple Inc., Cupertino, CA 95014, USA (e-mail: kirknie@gmail.com).

B. M. Hochwald is with the Department of Electrical Engineering, University of Notre Dame, Notre Dame, IN 46556, USA (e-mail: bhochwald@nd.edu).

This work was supported, in part, by NSF grants CCF-1403458 and ECCS-1509188.

into account is used to derive bounds on the ability to match multiple sources and loads over all $\omega$ with a lossless network. We expand this definition to include lossy networks.

Of particular interest is the application to loads that are closely-spaced antennas, such as may be found in multiple-input multiple-output (MIMO) communication systems. The "densification" of portable wireless communication devices, including cellular telephones, with multiple transmitter and receiver chains in close proximity, makes coupling difficult to avoid. Furthermore, there are situations where there are more antennas than RF chains [11]. Our analysis methods quantify the bandwidth attainable in the transmitters of these MIMO systems, where the RF amplifiers are treated as sources and the coupled antennas are the loads. Part II, in particular, shows how realistic antennas are modeled to obtain accurate bandwidth results.

We consider an RF system where $M$ sources drive $N$ loads through an arbitrary passive $(M + N)$-port matching network. The $M$ input ports on the multiport network connect to the sources, and $N$ output ports connect to the loads. No relationship between $M$ and $N$ is assumed. Our goal is to transfer maximum power from sources with known characteristic impedances to loads with known frequency-dependent impedances. The loads are dissipative and potentially non-reciprocal; the network can be lossy and also non-reciprocal. The quality of match between the sources and the loads at a frequency $\omega$ is then determined by the power lost either because it is returned to the sources or because it is dissipated in the network. We derive and utilize bounds on this quality metric when an arbitrary passive $(M + N)$-port network is used. Our bounds can readily be calculated from the frequency-dependent S-matrix of the loads.

We do not consider noise in our analysis, and hence our methods apply more to transmitters driving loads than to receivers, where the noise figures of the input amplifiers play an important role. We are interested in the "input bandwidth" of the loads and do not evaluate the ability of the loads to use their input power efficiently.

When the loads are coupled, there is only a nominal association of source $i$ with load $i$ for $i = 1, \ldots, N$ since source $i$ potentially also stimulates load $j \neq i$ if the two loads are coupled. In minimizing the amount of lost power, a matching network between the sources and loads therefore could connect any source with any of the loads, and conversely; examples include decoupling networks [12]–[15], which ensure all power from the sources is delivered to the loads at a design frequency. An effective network prevents reflection by decoupling the loads from each other over as wide a frequency range as possible. Generally, for a given design frequency $\omega_d$, there are frequencies $\omega_1 < \omega_d$ and $\omega_2 > \omega_d$, where the fractional power delivered to the loads falls below some prescribed threshold. The larger $\omega_2 - \omega_1$ is, the larger the bandwidth.

Techniques involving active non-Foster circuits [16], [17] and adaptive matching [18]–[21] are not discussed. Non-Foster circuits such as those used to realize negative capacitance and inductance values can theoretically cancel the reactive Foster behavior in the loads, and achieve an impedance match between the sources and the loads over a wider frequency band than passive networks. Adaptive matching uses tunable capacitors and inductors to match antennas whose impedance is affected by the environment or carrier frequency.

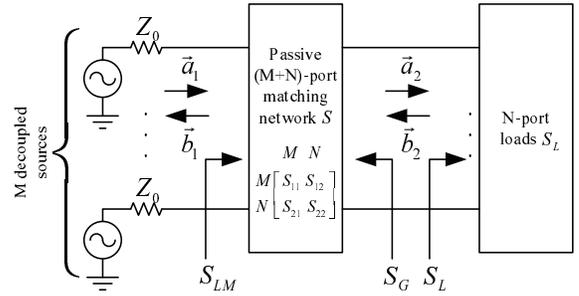

Fig. 1. An RF system where $M$ uncorrelated sources drive $N$ loads having S-matrix $S_L$ through a passive $(M+N)$-port matching network $S$ (the complex-frequency argument $s$ is omitted). We use $S_{LM}$ to denote the S-matrix as seen at the input of the network, and $S_G = S_{22}$ to denote the S-matrix as seen at the output.

The basic model assumptions are given in Section II. The definitions and notations we use are similar to those in [9]. The principal bounds we use are presented in Section III. Details on the mechanics of how to apply these bounds are provided within the subsections of Section III. Analytical applications of the results appear in Section IV, where scaling laws are derived and network architectures are analyzed. Section VI concludes.

The results presented herein assume the loads can be accurately modeled by rational functions of frequency. Since not all loads are necessarily rational, Part II examines how rational functions can be used as approximations for arbitrary loads. We present examples of how to use the bounds in practice. Proofs of all the results are also given.

## II. PROBLEM DEFINITION AND NOTATIONS

### A. System description

Figure 1 shows an RF system where $N$ dissipative loads are driven by $M$ decoupled sources with real impedance $Z_0$, the characteristic impedance of the system. Let $S_L(s)$ be the $N \times N$ S-matrix of the dissipative loads as a function of the complex frequency $s = \sigma + j\omega$, where $\sigma$ and $\omega$ are real. $S_L(s)$ is obtained by extending the S-matrix of the loads $S_L(j\omega)$ as a function of the angular frequency $\omega$ to the whole complex plane (WCP). Mathematically, $S_L(s)$ can be thought of as the transfer function between the $N \times 1$ incident and reflected waves $\vec{a}_2 e^{st}$ and $\vec{b}_2 e^{st}$, where $t$ represents time. Therefore, we have $\vec{b}_2(s) = S_L(s)\vec{a}_2(s)$. If the loads are reciprocal then $S_L(s)$ is symmetric. If the loads are coupled, at least one off-diagonal entry of $S_L(s)$ is non-zero. The impedance matrix $Z_L(j\omega)$ of the loads can be obtained by $Z_L(j\omega) = Z_0(I + S_L(j\omega))(I - S_L(j\omega))^{-1}$.

The $M$ sources and $N$ loads are matched by inserting a passive $(M + N)$-port network between them, as indicated in Figure 1. The $M$ input ports of the network are connected to the sources, and the $N$ output ports are connected to the loads. The network is not necessarily lossless or reciprocal, so we allow standard passive capacitive and inductive elements as well as non-reciprocal ferromagnetic components in its design.

Let $S(s)$ be the $(M + N) \times (M + N)$ S-matrix of the multiport network as a function of $s$, partitioned as in Figure 1

$$S(s) = \begin{array}{cc} & \begin{array}{cc} M & N \end{array} \\ \begin{array}{c} M \\ N \end{array} & \begin{pmatrix} S_{11}(s) & S_{12}(s) \\ S_{21}(s) & S_{22}(s) \end{pmatrix}. \end{array} \tag{1}$$

Let $S_{LM}(s)$ denote the $M \times M$ S-matrix of the cascade of the network and the loads, and $S_G(s)$ be the $N \times N$ S-matrix seen from the output ports of the network. Since the input to the network is terminated by sources with characteristic impedance $Z_0$, it follows that $S_G(s) = S_{22}(s)$, and

$$S_{LM}(s) = S_{11}(s) + S_{12}(s)S_L(s)(I - S_G(s)S_L(s))^{-1}S_{21}(s). \tag{2}$$

As in Figure 1, let $M \times 1$ vector $\vec{a}_1(s)$ be the incident wave to the network. Then the reflected wave is $\vec{b}_1(s) = S_{LM}(s)\vec{a}_1(s)$. Since the network is potentially lossy, the incident power may be reflected to the sources (return loss), or dissipated in the network (insertion loss). We now define a measure of performance that includes both of these effects.

### B. Power loss ratio

Our measure of performance of a matching network between the sources and loads is given by the fraction of source power that is not transferred to the loads, as a function of frequency. At frequency $s = j\omega$, the total instantaneous power from the $M$ sources is $\|\vec{a}_1(j\omega)\|^2$, and the total instantaneous power delivered to the $N$ loads is $\|\vec{a}_2(j\omega)\|^2 - \|\vec{b}_2(j\omega)\|^2 \geq 0$. The power lost due to dissipation and reflection is therefore $\|\vec{a}_1(j\omega)\|^2 - (\|\vec{a}_2(j\omega)\|^2 - \|\vec{b}_2(j\omega)\|^2)$.

*Definition 1:* The *power loss ratio* at frequency $j\omega$ is the ratio between the expected power loss and the expected total incident power at $j\omega$:

$$r^2(\omega) = \frac{\mathbb{E}\left[\|\vec{a}_1(j\omega)\|^2 - (\|\vec{a}_2(j\omega)\|^2 - \|\vec{b}_2(j\omega)\|^2)\right]}{\mathbb{E}\|\vec{a}_1(j\omega)\|^2}, \tag{3}$$

where the expectation is over the random input signals.

By convention, when we use $r(\omega)$ we mean the positive square root of (3), and by construction, $0 \leq r(\omega) \leq 1$ where values close to zero indicate that little source power is being lost and therefore most of it is being delivered to the loads. Values close to one indicate most of the power is lost to dissipation or reflection. We note that $r(\omega) = 0$ means that the loads are perfectly matched and decoupled from one another. When the matching network $S(s)$ is lossless, the power loss ratio is equivalent to the power reflection ratio defined in [9]. By considering power dissipation in the network in addition to power reflection by the network, we handle situations that appear favorable because the reflected energy is low, but are actually unfavorable because the network (rather than the load) is dissipating the incident power. This issue is raised in [22].

### C. Experimental measurement of $r(\omega)$

Because (3) includes an expectation, we devote a few words to the measurement of $r(\omega)$ in practice. The incident signals from the sources are assumed to have independent random phases and instantaneous powers at each frequency, and the expectation is taken over all such uncorrelated incident signals. Nevertheless, the expectation is unnecessary when $M = 1$ since the amplitude and phase of a single source does not affect the power loss ratio. In this case, $1 - r^2(\omega)$ is equivalent to the standard transducer power gain [8], [23], and the expectations in the numerator and denominator of (3) may be dropped.

When $M > 1$ the expectations play the important role of averaging over all amplitude and phase combinations of the input signals. Its importance can be demonstrated by examining $M = 2$, where it is well known that even-mode (in-phase) and odd-mode (out-of-phase) signals can elicit very different reflective responses from a two-port system. Since the sources are uncorrelated, no preference is given to even or odd mode signals, and the expectation eliminates mode dependence by averaging over both of them. Hence, $r(\omega)$ can be thought of as an "average loss" experienced when the loads are stimulated by uncorrelated sources.

The value of $r^2(\omega)$ can be experimentally measured by forming the sample average of the numerator $\|\vec{a}_1(j\omega)\|^2 - (\|\vec{a}_2(j\omega)\|^2 - \|\vec{b}_2(j\omega)\|^2)$ for a variety of inputs $\vec{a}_1(j\omega)$, and taking the ratio of this average to the sample average of the denominator $\|\vec{a}_1(j\omega)\|^2$. This ratio of averages converges to $r^2(\omega)$ as more measurements are taken with all possible source phase and amplitude combinations.

### D. Definition of bandwidth

The network $S(s)$ should be constructed to make $r(\omega)$ as small as possible over a prescribed bandwidth, or make the bandwidth as wide as possible for a prescribed threshold. Usually, bandwidth is measured in the vicinity of a design frequency, which we denote as $\omega_d$. A decoupling network [15] enforces $r(\omega) = 0$ at $\omega = \omega_d$. We can define the bandwidth of the combined network and loads using (3).

*Definition 2:* The *bandwidth* is the frequency range for which $r(\omega)$ is no greater than a threshold $\tau > 0$ in the vicinity of a design frequency $\omega_d$:

$$\omega_{\text{BW}}(\tau, \omega_d) = \max_{\substack{\omega_1 \leq \omega_d \leq \omega_2 \\ r(\omega) \leq \tau, \forall \omega \in [\omega_1, \omega_2]}} \omega_2 - \omega_1. \tag{4}$$

Let the elements of $\vec{a}_1(j\omega)$, representing the incident signal from $M$ decoupled sources at frequency $j\omega$, have equal expected square-magnitude over all frequency, and have uniformly distributed random phases in $[0, 2\pi)$ that are independent of the amplitudes and each other. Then (3) yields

$$r^2(\omega) = 1 - \frac{\mathbb{E}\,\text{tr}\{\vec{a}_2^H(j\omega)(I - S_L^H(j\omega)S_L(j\omega))\vec{a}_2(j\omega)\}}{\mathbb{E}\,\text{tr}\{\vec{a}_1^H(j\omega)\vec{a}_1(j\omega)\}}.$$

Because the phases are independent and uniformly distributed, $\mathbb{E}[\vec{a}_1(j\omega)\vec{a}_1^H(j\omega)]$ is a multiple of the identity matrix. We apply $\vec{a}_2(j\omega) = (I - S_G(j\omega)S_L(j\omega))^{-1}S_{21}(j\omega)\vec{a}_1(j\omega)$ to obtain

$$r^2(\omega) = 1 -$$

$$\frac{\mathrm{tr}\{S_{21}^H(I-S_GS_L)^{-H}(I-S_L^HS_L)(I-S_GS_L)^{-1}S_{21}\}}{M}, \quad (5)$$

where $\mathrm{tr}(\cdot)$ denotes trace, $^H$ denotes Hermitian transpose; the frequency argument $j\omega$ is omitted on the right-hand side of (5).

When $M > N$, $r^2(\omega) \geq 1 - N/M$ since the matrix inside the trace on the right-hand side of (5) is a rank-$N$ positive semidefinite matrix whose eigenvalues are less than or equal to one. Hence, a certain fraction of the source power is always reflected or absorbed, and letting $\tau < \sqrt{1 - N/M}$ in (4) always obtains zero bandwidth. We therefore assume $\sqrt{1 - N/M} < \tau < 1$ for $M > N$; for $M \leq N$, we have $0 < \tau < 1$.

Note that "ganging" amplifiers through couplers to attain high output power with a single load appears to be an example where $M > N$. But such ganged sources are correlated since each carries the same signal, possibly differing only in a constant relative phase or amplitude. Since we assume uncorrelated sources, ganged amplifiers (and other correlated sources) should be considered as a single source to apply our results.

Although the incident signals from the sources are assumed to be uncorrelated, the output of the matching network will have correlated components when $M < N$ since any network driving all $N$ loads will necessarily derive its signals from the $M$ sources. As an example, a beamformer can be modeled as $M = 1$ source driving $N > 1$ loads in a fixed phase relationship. We consider beamforming in Section V-A.

### E. Properties of S-matrices

We briefly summarize some properties of S-matrices since they play an important role. Passive real networks have S-matrices that are real-rational, Hurwitzian, and bounded; this applies, for example, to $S_L(s)$ and $S(s)$. The definitions of these terms can be found in [23]–[25]. We also employ definitions of poles and zeros of rational matrices that are widely used in multivariable control theory [26]. For an arbitrary rational matrix $A(s)$:

- **Poles:** are the roots of the pole polynomial of $A(s)$, where this polynomial is the monic least common multiple of the denominators of all minors of all dimensions of $A(s)$.
- **Zeros:** are the roots of the zero polynomial of $A(s)$, where this polynomial is the monic greatest common divisor of the numerators of all minors of dimension $L$, and $L$ is the normal rank of $A(s)$. It is assumed that these minors have the pole polynomial as their denominators.

The normal rank of a matrix is its maximum rank among all $s \in \mathbb{C}$; the use of "normal" refers to the rank almost everywhere in the complex plane [24, A.70]. We use LHP to denote the left-half complex plane ($\mathrm{Re}\{s\} < 0$) and RHP to denote the right-half ($\mathrm{Re}\{s\} > 0$). Throughout, we use $p_{L,i}$ and $z_{L,i}$, $i = 1, 2, \ldots$ to represent the poles and zeros over the WCP (whole complex plane) of $S_L(s)$. Since $S_L(s)$ is Hurwitzian, it has no RHP poles.

We assume $I - S_L^T(-s)S_L(s)$ is full normal rank. This is satisfied if the loads are strictly dissipative, so that there is no combination of load signals that is completely reflected for all

$s$. On the other hand, we also assume that there exists an $s_0$ with $\mathrm{Re}\{s_0\} \geq 0$ such that

$$S_L^T(-s_0)S_L(s_0) = I. \quad (6)$$

In general, $s_0$ is arbitrary and can be infinite.

When $s_0$ is purely imaginary $s_0 = j\omega_0$, (6) has the interpretation of modeling the loads as perfect reflectors at frequency $\omega_0$ since $S_L^T(-j\omega) = S_L^H(j\omega)$ and therefore the singular values of $S_L(j\omega_0)$ are all unity. Because $S_L(s)$ is a bounded matrix, (6) is equivalent to $|\det S_L(j\omega_0)| = 1$.

When $s_0$ has positive real part, the physical interpretation of (6) is elusive but we still use the terminology "perfect reflector" at $s_0$. Examples of load structures with such an $s_0$ are given in Section IV-B2.

For any $s_0$, (6) must also hold if we replace $s_0$ with $-s_0$. Therefore, without loss of generality, we only consider $\mathrm{Re}\{s_0\} \geq 0$. If there are multiple distinct values of $s_0$ for which (6) holds then the bounds presented herein apply to each value separately. We therefore consider only a single distinct $s_0$.

### III. BROADBAND MATCHING BOUNDS

We present the principal bounds used in this paper. They are derived in detail in Part II, but knowledge of the derivations is not needed to apply the bounds. A description of how to use these bounds appears in Section III-B. Conditions for achieving equality are presented in Section III-C. The bounds depend on zeros and poles of $S_L(s)$, and techniques to obtain these are discussed in Section III-D and Part II.

### A. Principal bounds

**Bound 1:** For $s_0 = j\omega_0$,

$$\int_0^\infty \frac{(\omega_0 - \omega)^{-2} + (\omega_0 + \omega)^{-2}}{2} \log \frac{1}{r(\omega)} d\omega$$
$$\leq \frac{-\pi}{2M} \left[ \sum_i (p_{L,i} - j\omega_0)^{-1} + \sum_i (z_{L,i} + j\omega_0)^{-1} \right]. \quad (7)$$

**Bound 2:** For $\mathrm{Re}\{s_0\} > 0$,

$$\int_0^\infty \frac{\mathrm{Re}[(s_0 - j\omega)^{-1} + (s_0 + j\omega)^{-1}]}{2} \log \frac{1}{r(\omega)} d\omega$$
$$\leq \frac{-\pi}{2M} \log \left| \det S_L(s_0) \cdot \frac{\prod_i (s_0 + z_{L,i})}{\prod_i (s_0 - z_{L,i})} \right|. \quad (8)$$

**Bound 3:** For $s_0 = \infty$,

$$\int_0^\infty \log \frac{1}{r(\omega)} d\omega \leq \frac{-\pi}{2M} \left( \sum_i p_{L,i} + \sum_i z_{L,i} \right). \quad (9)$$

Bound 1 is useful for loads that are modeled as electrically open or short at some frequency $j\omega_0$. For example, some antennas are capacitive relative to ground when they are "electrically small" compared to the signal wavelength. Thus, they are effectively an open circuit at $s_0 = 0$; equivalently $S_L(0) = I$.

Bound 2 applies to loads that are modeled as a mixture of resistive and reactive components [27]. An example of this is given in Section IV-B2.

| $s_0 = j\omega_0$ | $f(\omega) = \frac{1}{2}[(\omega_0 - \omega)^{-2} + (\omega_0 + \omega)^{-2}]$ |
| | $B = \frac{-\pi}{2M}\left[\sum_i (p_{L,i} - j\omega_0)^{-1} + \sum_i (z_{L,i} + j\omega_0)^{-1}\right]$ |
| $\mathrm{Re}\{s_0\} > 0$ | $f(\omega) = \frac{1}{2}\mathrm{Re}[(s_0 - j\omega)^{-1} + (s_0 + j\omega)^{-1}]$ |
| | $B = \frac{-\pi}{2M}\log\left|\det S_L(s_0) \cdot \frac{\prod_i(s_0 + z_{L,i})}{\prod_i(s_0 - z_{L,i})}\right|$ |
| $s_0 = \infty$ | $f(\omega) = 1$ |
| | $B = \frac{-\pi}{2M}\left(\sum_i p_{L,i} + \sum_i z_{L,i}\right)$ |

Bound 3 applies to loads that are modeled as open or short circuits at infinite frequency. The classical model of a parallel resistive and capacitive load used to demonstrate the Bode-Fano bound falls into this category. A bound similar to (9) is presented in [9]; however, the bound in [9] requires $M = N$ and the matching network to be lossless.

In all three cases, the bounds have the form

$$\int_0^\infty f(\omega)\log\frac{1}{r(\omega)}\,d\omega \le B, \tag{10}$$

where $f(\omega) \ge 0$. The form of $f(\omega)$ depends on the location of $s_0$, and the computation of $B$ depends also on the poles and zeros of $S_L(s)$. Because $0 \le r(\omega) \le 1$, we have $\log(1/r(\omega)) \ge 0$. Hence, $f(\omega)\log(1/r(\omega)) \ge 0$ for any $\omega$. Clearly, $B$ must be positive as well. We use (10) to generically indicate any of (7)–(9). The number of sources appears as $M$ in the denominators of all three bounds. The forms of $f(\omega)$ and $B$ are summarized in Table I.

### B. How to use bounds

Suppose that we wish to assess the achievable bandwidth $[\omega_1, \omega_2]$ of a set of loads, where $\omega_1 < \omega_2$. Our measure of achievability is that for some threshold $\tau > 0$, the overall system should obey $r(\omega) \le \tau$ for $\omega \in [\omega_1, \omega_2]$. Hence the combined multiport network and loads reflects (or absorbs) no more than $\tau$ within the passband. We assume that $S_L(s)$ is available to us (we have more to say about this in Section III-D) and we would like to know $\omega_{\mathrm{BW}}(\tau, \omega_d)$ defined in (4).

Suppose the loads obey $S_L^T(0)S_L(0) = I$ so that Bound 1 (first row of Table I) with $\omega_0 = 0$ applies to any passive network used for these loads. Let the right-hand side of (7) be denoted $B_1 > 0$, which depends only on $S_L(s)$. Then

$$B_1 \ge \int_0^\infty \omega^{-2}\log\frac{1}{r(\omega)}\,d\omega \ge \log\frac{1}{\tau}\int_{\omega_1}^{\omega_2}\omega^{-2}\,d\omega$$
$$= \log\frac{1}{\tau}\left(\frac{1}{\omega_1} - \frac{1}{\omega_2}\right).$$

The first inequality applies to any network, while the second inequality applies to a network with the desired passband characteristics. Hence,

$$\frac{1}{\omega_1} - \frac{1}{\omega_2} \le \frac{B_1}{\log(1/\tau)}. \tag{11}$$

Clearly, this inequality imposes a constraint on $(\omega_1, \omega_2)$ pairs.

Bounds 2 and 3 can be applied in a similar fashion. If $S_L^H(\infty)S_L(\infty) = I$, (9) gives

$$\omega_2 - \omega_1 \le \frac{B_3}{\log(1/\tau)}, \tag{12}$$

where $B_3$ is the right-hand side of (9). This gives us a direct bound on the bandwidth $\omega_{\mathrm{BW}}(\tau, \omega_d)$ achievable for all $\omega_d$. Equations (11) and (12) are complementary in that both can be in force simultaneously.

We now discuss conditions under which these bounds can be achieved.

### C. Conditions for equality

We distinguish between conditions for equality in (7)–(9) and conditions for equality in (11), (12). The former apply to any passive real network and the latter apply to networks with particular passband characteristics. Bounds (7)–(9) have a common set of conditions for achieving equality:

1) $S_{21}(s)S_{21}^T(-s) + S_G(s)S_G^T(-s) = I$ for all $s$
2) The $M \times M$ matrix

$$S_{21}^H(I - S_G S_L)^{-H}(I - S_L^H S_L)(I - S_G S_L)^{-1}S_{21} \tag{13}$$

has equal singular values for all $s = j\omega$
3) $I - S_L(s_0)S_G(s_0)$ is non-singular
4) $S_L^T(-s) - S_G(s)$ has no zeros in the RHP

where $s_0$ is defined in (6). These four conditions correspond to four possible impediments to achieving the bounds. Meeting all the conditions is sufficient to attain equality in the bounds, but Conditions 1, 2 and 4 are also necessary. The $N \times N$ matrix $S_G(s)$ (see Figure 1) plays a prominent role in these conditions and can be readily measured or modeled by connecting pairs of output ports of the matching network to a network analyzer while terminating its remaining ports with characteristic impedances.

These conditions have physical interpretations. Condition 1 is satisfied for lossless networks since we are guaranteed that $S_{21}(s)S_{21}^T(-s) + S_{22}(s)S_{22}^T(-s) = I$ for all $s$ because $S(s)$ is a para-unitary matrix, and $S_G(s) = S_{22}(s)$.

Condition 2 requires the singular values of (13) to be equal for all $s = j\omega$. When this is satisfied, the total dissipated power at the loads depends only on the total incident power $\|\vec{a}_1(j\omega)\|^2$, and not the "direction" of $\vec{a}_1(j\omega)$.

Condition 3 is a "non-degenerate" condition that we illustrate with an example: Suppose the loads are capacitive to ground, and hence are reflective with $S_L(\infty) = -I$. If the output impedance of the matching network is also capacitive, then $S_G(\infty) = -I$ and Condition 3 is violated. Hence, in this example, a matching network that wants to achieve high bandwidth should avoid capacitive output impedance. It turns out that Condition 3 is superfluous in Bound 2 because $S_L(s)$ and $S_G(s)$ are bounded matrices; hence $S_L(s)S_G(s)$ is also bounded and $I - S_L(s_0)S_G(s_0)$ is always non-singular for $\mathrm{Re}\{s_0\} > 0$ [24, 7.22]. Additional details on Condition 3 can be found in [9].

Condition 4 is a "minimum-phase" condition since the RHP zeros of $S_L^T(-s) - S_G(s)$ are the same as the RHP zeros of $S_{GM}(s)$, which involves a "Darlington equivalent" network representation of the loads [28], [29] as described in Part II. However, knowledge of the Darlington equivalent network is not needed to check this condition. We have more to say about network architectures that cannot meet this condition in Section IV-C.

Assuming Conditions 1–4 are met by the matching network, we obtain equality in (11), (12) if the network also achieves the ideal response $r(\omega) = \tau$ for $\omega \in [\omega_1, \omega_2]$ and $r(\omega) = 1$ elsewhere. Any network such that $r(\omega) < 1$ for $\omega \notin [\omega_1, \omega_2]$ sustains a non-negative "shaping loss" which is the difference between the integral over all $\omega$ of the left-hand sides of (7)–(9) versus the integral over $[\omega_1, \omega_2]$:

$$\text{shaping loss} =$$
$$\int_0^\infty f(\omega) \log \frac{1}{r(\omega)} d\omega - \int_{\omega_1}^{\omega_2} f(\omega) \log \frac{1}{r(\omega)} d\omega. \quad (14)$$

For a network whose shaping loss is positive, some performance is lost in the band of interest because either $\omega_2 - \omega_1$ could potentially be made larger, or $\tau$ could be made smaller, by redesigning the network to have larger $r(\omega)$ for $\omega \notin [\omega_1, \omega_2]$. We provide an example of the shaping loss computation in Part II, Section III-A.

### D. How to obtain $S_L(s)$ and its poles and zeros

If analytical circuit models for the loads are known, $S_L(s)$ is uniquely determined by the standard Laplace transform representations of the model impedance or admittance matrix and using the formula $S_L(s) = (Z_L(s) + Z_0 I)^{-1}(Z_L(s) - Z_0 I)$. Examples of this are presented in Section IV-B.

Absent an analytical model, numerical methods that model the $S_L(s)$ of the loads are needed to extract its poles and zeros. The modeling methods are discussed and illustrated by example in Part II.

## IV. Bandwidth Analysis

The bounds yield a variety of conclusions when they are applied to various system configurations. We first examine decoupled loads in Section IV-A and then coupled loads in IV-B. We then identify network architectures that can achieve the bounds in Section IV-C.

### A. Decoupled loads

Let $S_L(s) = S_l(s)I$ where $S_l(s)$ is a scalar and $I$ is an $N \times N$ identity matrix. Thus, there are $N$ decoupled identical loads. Let $S_l(s)$ satisfy (6) for some $s_0$, and let $B_l$ be computed using Table I applied to $S_l(s)$ for $M = 1$. Then $S_L(s)$ satisfies (6) at the same $s_0$, and has poles and zeros at the same locations as $S_l(s)$, each with multiplicity $N$. It follows from Table I that the bound for $N$ loads is $N$ times the bound for a single load. We have proven the following theorem.

*Theorem 1:* For $N$ identical decoupled loads driven by $M$ sources

$$\int_0^\infty f(\omega) \log \frac{1}{r(\omega)} d\omega \le \frac{N}{M} B_l. \quad (15)$$

If $N = M$, (15) is simply $B_l$. This is not surprising because we have assumed the sources and loads are decoupled and therefore we are essentially examining $N$ identical isolated systems, each with bound $B_l$. However, (15) scales linearly with $N$ for a fixed $M$. This formula suggests that $N$ identical decoupled loads can achieve $N$ times the bandwidth of one load as long as the matching network is designed properly.

Matching network architectures can be thought of in terms of their ability to achieve linear-in-$N$ bandwidth behavior when used with a set of loads with linear-in-$N$ behavior. In Section IV-C we show that certain network architectures provably have linear-in-$N$ behavior, while others do not. As shown in the next section, this linear-in-$N$ behavior extends to coupled loads under some conditions.

### B. Circulantly-coupled loads

Loads with circulant $S_L(s)$ are especially easy to manipulate mathematically because, while the eigenvalues of $S_L(s)$ depend on $s$, its eigenvectors do not. This makes computing $p_{L,i}$ and $z_{L,i}$ straightforward. Load structures that lead to symmetric circulant $S_L(s)$ are identical, and the coupling between load $i$ and neighbor $j$ depends on $\min\{|i - j|, N - |i - j|\}$.

Two identical loads automatically have circulant symmetry, as long as they have reciprocal coupling. Three or more loads can be placed in a circular or spherical arrangement to yield circulant $S_L(s)$.

*1) Loads exemplifying Bound 1:* Figure 2(a) illustrates $N$ circulantly-coupled loads. The loads consist of an $N$-port inductive network with impedance matrix $s Z_l$ and an $N$-port series LC network with impedance matrix $\frac{s^2/\omega_0^2 + 1}{2s} Z_{lc}$ resonating at $j\omega_0$. These networks are connected in parallel to each other, and terminated by isolated characteristic impedances $Z_0$. The parallel $N$-port networks are shown in Figure 2(b,c); the inductive network has each port grounded through $L_0$ and every pair of ports $i$ and $j$ is connected through $L_\ell$, where $\ell = \min\{|i - j|, N - |i - j|\}$ is the "distance" between ports $i$ and $j$. The series LC network has each port grounded through $L_0'$ and $C_0'$, and every pair of ports $i$ and $j$ is connected through $L_\ell'$ and $C_\ell'$, where $\sqrt{L_0' C_0'} = \sqrt{L_\ell' C_\ell'} = 1/\omega_0^2$. The $N \times N$ matrices $Z_l$ and $Z_{lc}$ are circulant symmetric.

Since circulant matrices have eigenvectors that are columns of a discrete Fourier transform (DFT) matrix, we obtain the following eigenvalue decompositions:

$$Z_l = W \Lambda_l W^H, \quad Z_{lc} = W \Lambda_{lc} W^H,$$

where

$$W = \frac{1}{\sqrt{N}} \begin{bmatrix} 1 & 1 & \cdots & 1 \\ 1 & e^{-\frac{j 2\pi}{N}} & \cdots & e^{-\frac{j 2\pi (N-1)}{N}} \\ \vdots & \vdots & \ddots & \vdots \\ 1 & e^{-\frac{j 2\pi (N-1)}{N}} & \cdots & e^{-\frac{j 2\pi (N-1)^2}{N}} \end{bmatrix} \quad (16)$$

is the $N \times N$ unitary DFT matrix, and $\Lambda_l$ and $\Lambda_{lc}$ are real positive diagonal matrices representing the eigenvalues of $Z_l$ and $Z_{lc}$, respectively.

The impedance matrix of the loads is

$$Z_L(s) = \left( \frac{1}{s} Z_l^{-1} + \frac{2s}{s^2/\omega_0^2 + 1} Z_{lc}^{-1} + \frac{1}{Z_0} I \right)^{-1}.$$

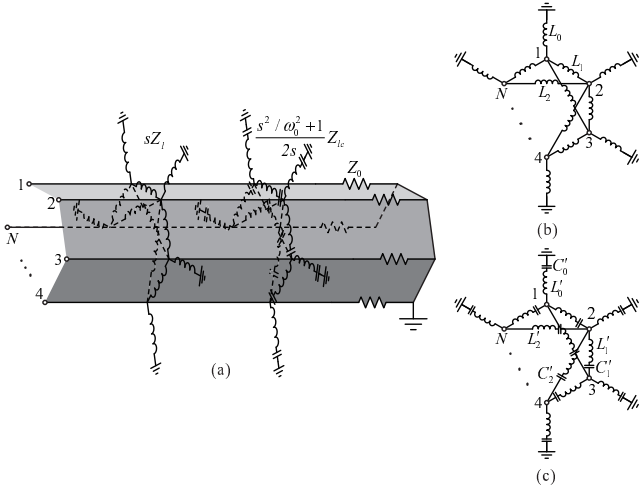

Fig. 2. (a) $N$ circulantly-coupled loads consisting of an inductive network and an LC network in parallel, terminated in a set of characteristic impedances $Z_0$. The inductive and the LC portions of the network are shown in (b,c), and have impedance matrices $sZ_l$ and $\frac{s^2/\omega_0^2+1}{2s}Z_{lc}$, respectively. The loads are coupled to ground through $L_0$ and series $L_0'$ and $C_0'$, and every load pair is coupled through $L_\ell$ and series $L_\ell'$ and $C_\ell'$, where $\ell$ is the "distance" between the pair.

Then $S_L(s)$ can be obtained from $S_L(s) = (Z_L(s) + Z_0 I)^{-1}(Z_L(s) - Z_0 I)$, which is

$$S_L(s) = W \frac{-Z_0[(2\Lambda_{lc}^{-1}\omega_0^2 + \Lambda_l^{-1})s^2 + \Lambda_l^{-1}\omega_0^2]}{2s^3 + Z_0(2\Lambda_{lc}^{-1}\omega_0^2 + \Lambda_l^{-1})s^2 + 2\omega_0^2 s + Z_0\Lambda_l^{-1}\omega_0^2} W^H. \quad (17)$$

Note, we write $(\cdot)^{-1}$ as $\frac{1}{(\cdot)}$ when we take inverses of diagonal matrices. Because $S_L(s)$ is a circulant matrix, only its eigenvalues depend on $s$ and the poles and zeros of $S_L(s)$ are therefore the poles and zeros of the individual eigenvalues.

All $N$ loads present short circuits to ground at $s_0 = 0$ and $j\omega_0$, where the loads become perfect reflectors. Hence, $S_L(0) = S_L(j\omega_0) = -I$, and (6) is satisfied. We can then obtain two distinct bounds by applying Bound 1 for $s_0 = 0$ and $j\omega_0$. Using $p_{L,i}, z_{L,i}$ calculated from (17), Bound 1 yields

$$\int_0^\infty \omega^{-2} \log \frac{1}{r(\omega)} d\omega \leq \frac{-\pi}{2M} \sum_i \frac{-2\lambda_{l,i}}{Z_0} = \frac{\pi \cdot \operatorname{tr} Z_l}{MZ_0}, \quad (18a)$$

$$\int_0^\infty \frac{(\omega_0 - \omega)^{-2} + (\omega_0 + \omega)^{-2}}{2} \log \frac{1}{r(\omega)} d\omega$$
$$\leq \frac{-\pi}{2M} \sum_i \frac{-2\lambda_{lc,i}}{Z_0\omega_0^2} = \frac{2\pi \cdot \operatorname{tr} Z_{lc}}{MZ_0\omega_0^2}. \quad (18b)$$

The $i$th diagonal element of $Z_l$ represents the inductance relative to ground of port $i$ of the inductive network in Figure 2(b), measured with the remaining ports open. Because $Z_l$ is circulant its diagonal elements are all equal. Hence $(1/N)\operatorname{tr} Z_l = L_{\text{eq},N}$, where $L_{\text{eq},N}$ is the inductance of any port relative to ground. A similar conclusion holds for the LC network in Figure 2(c). Let the equivalent series LC branch of any port relative to ground have inductance $L_{\text{eq},N}'$ and capacitance $C_{\text{eq},N}'$, where $L_{\text{eq},N}' C_{\text{eq},N}' = 1/\omega_0^2$. Then $(1/N)\operatorname{tr} Z_{lc} = \omega_0^2 L_{\text{eq},N}' + 1/C_{\text{eq},N}'$. We can rewrite (18) as

$$\int_0^\infty \omega^{-2} \log \frac{1}{r(\omega)} d\omega \leq \frac{N\pi L_{\text{eq},N}}{MZ_0}, \quad (19a)$$

$$\int_0^\infty \frac{(\omega_0 - \omega)^{-2} + (\omega_0 + \omega)^{-2}}{2} \log \frac{1}{r(\omega)} d\omega$$
$$\leq \frac{N\pi}{MZ_0} \left( L_{\text{eq},N}' + \frac{1}{\omega_0^2 C_{\text{eq},N}'} \right), \quad (19b)$$

Let $L_{\text{eq},N}$, $L_{\text{eq},N}'$ and $C_{\text{eq},N}'$ approach respective limits $L_{\text{eq}}$, $L_{\text{eq}}'$ and $C_{\text{eq}}'$ as $N \to \infty$. We have proven the following theorem.

*Theorem 2:* The linear-in-$N$ behavior shown in Theorem 1 for decoupled loads extends to circulantly-coupled loads provided $L_{\text{eq}} > 0$ and $L_{\text{eq}}' > 0$ (or $C_{\text{eq}}' < \infty$).

The conditions $L_{\text{eq}} > 0$ and $L_{\text{eq}}' > 0$ (or $C_{\text{eq}}' < \infty$) are equivalent to ensuring that the inductance (or capacitance) of any port relative to ground does not go to zero (or infinity) as $N \to \infty$, and hence there are not "too many" parallel paths to ground from any port. This is equivalent to imposing a condition that the coupling between loads not be "too strong".

For a given $N$ and $M$, the bounds in (19) for Figure 2 increase as inductive and capacitive coupling components are removed, since $L_{\text{eq},N}$ increases and approaches $L_0$ as all cross-inductive elements are removed; similarly $L_{\text{eq},N}'$ increases to $L_0'$ and $C_{\text{eq},N}'$ decreases to $C_0'$ because $L_{\text{eq},N}$, $L_{\text{eq},N}'$ and $C_{\text{eq},N}'$ are obtained as $L_0$, $L_0'$ and $C_0'$ in parallel with the remaining parts of the networks.

However, we cannot conclude that coupling always has a negative effect on bandwidth. In fact, Part II of this paper shows that coupling between dipole antennas has a non-monotonic effect on the bandwidth bound. This dichotomy in behavior between the model in Figure 2 and the dipoles in Part II is not contradictory because there is no requirement that dipoles should be modeled by fixed $L_0$, $L_0'$, and $C_0'$ as the coupling between them changes.

*2) Loads exemplifying Bound 2:* Figure 3 illustrates $N$ resistors $Z_0$ terminated with two parallel $N$-port networks: one is resistive with $N \times N$ circulant impedance matrix $Z_r$, and the other is capacitive with $N \times N$ circulant impedance matrix $\frac{1}{s}Z_c$. The impedance matrix of the loads $Z_L(s)$ is

$$Z_L(s) = Z_0 I + (Z_r^{-1} + sZ_c^{-1})^{-1}.$$

Let $\Lambda_r$ and $\Lambda_c$ be the eigenvalue matrices of $Z_r$ and $Z_c$. Then $S_L(s)$ is

$$S_L(s) = W \frac{\Lambda_r \Lambda_c}{2Z_0\Lambda_r s + 2Z_0\Lambda_c + \Lambda_r \Lambda_c} W^H \quad (20)$$

where $W$ is given in (16).

We note that

$$I - S_L^T(-s)S_L(s) = W \frac{-4Z_0^2\Lambda_r^2 s^2 + 4Z_0(Z_0 I + \Lambda_r)\Lambda_c^2}{-4Z_0^2\Lambda_r^2 s^2 + (2Z_0 I + \Lambda_r)^2\Lambda_c^2} W^H.$$

Let the component values satisfy $\frac{\Lambda_c \sqrt{Z_0(Z_0 I + \Lambda_r)}}{Z_0\Lambda_r} = \sigma_0 I$ for some $\sigma_0 > 0$. Then (6) is satisfied for $s_0 = \sigma_0$. We substitute $\Lambda_c = \frac{Z_0\Lambda_r \sigma_0}{\sqrt{Z_0(Z_0 I + \Lambda_r)}}$ into (20), and apply (8) to yield

$$\int_0^\infty \frac{\sigma_0}{\sigma_0^2 + \omega^2} \log \frac{1}{r(\omega)} d\omega$$

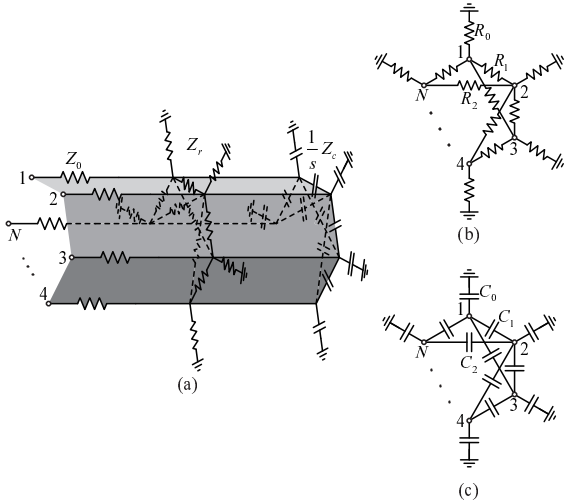

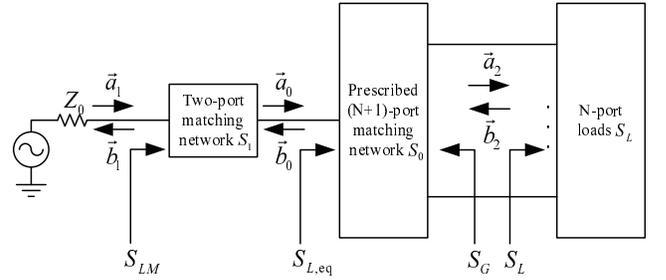

Fig. 4. An RF system where one source drives $N$ loads having S-matrix $S_L$ through a constrained matching network consisting of a prescribed $(N+1)$-port network $S_0$ and an arbitrary network $S_1$ (the complex-frequency argument $s$ is omitted). We use $S_{L,\text{eq}}$ to denote the S-parameter seen from the input of $S_0$.

Fig. 3. (a) $N$ circularly-coupled loads consisting of $N$ resistors $Z_0$ in series with a parallel resistive network $Z_r$ and a parallel capacitive network $\frac{1}{s}Z_c$. The resistive and the capacitive portions of the network are shown in (b,c), and have respective impedance matrices $Z_r$ and $\frac{1}{s}Z_c$. The loads are coupled to ground through $R_0$ and $C_0$, and every load pair is coupled through $R_\ell$ and $C_\ell$, where $\ell$ is the "distance" between the pair.

$$
\begin{aligned}
&\leq \frac{-\pi}{2M} \log \det \frac{\Lambda_r}{2Z_0 + \Lambda_r + 2\sqrt{Z_0(Z_0+\Lambda_r)}} \\
&= \frac{\pi}{2M} \sum_{i=1}^{N} \log \frac{2Z_0 + \lambda_{r,i} + 2\sqrt{Z_0(Z_0+\lambda_{r,i})}}{\lambda_{r,i}},
\end{aligned} \tag{21}
$$

where $\lambda_{r,i}$ are the diagonal elements of $\Lambda_r$. The linear-in-$N$ behavior of these loads depends on $\lambda_{r,1}, \ldots, \lambda_{r,N}$; we do not pursue this analysis any further here.

### C. Ability of network architecture to achieve bounds

Matching network architectures that cannot achieve equality in the bounds should be identified whenever possible and removed from consideration in large-bandwidth applications. The following theorem shows how $S_G(s)$ defined in Figure 1 can be compared with $S_L(s)$ to identify such networks. We thereby make Condition 4 in Section III-C more physically tangible.

*Theorem 3:* For matching networks satisfying Condition 3, if $S_G(s)$ has an eigenvector in common with $S_L(s)$, and the associated eigenvalue $\lambda_G(s)$ satisfies $|\lambda_G(j\omega)| = 1$ for all $\omega$, then Condition 4 cannot be satisfied and therefore the bounds cannot be achieved.

*Proof:* See the Appendix. ∎

Theorem 3 can be applied to the example of decoupled loads in Section IV-A. We assume $1 \leq M < N$ and Condition 3 is satisfied. Since $S_L(s) = S_l(s)I$, any vector is an eigenvector of $S_L(s)$. Theorem 3 then implies that any $S_G(s)$ that has $|\lambda_G(j\omega)| = 1$ cannot achieve equality in (15).

For example, networks where $S_G(s)$ is a real symmetric matrix cannot achieve equality. This follows because even if the network is lossless and Condition 1 is satisfied, implying that $S_{21}(s)S_{21}^T(-s) + S_G(s)S_G^T(-s) = I$ for all $s$, then because $S_{21}(s)$ is an $N \times M$ matrix and $S_G(s)$ is an $N \times N$ matrix, $S_G(s)$ has at least $N - M$ unit singular values. Since the moduli of the eigenvalues of a real symmetric matrix are equal to its singular values, at least one eigenvalue satisfies $|\lambda_G(j\omega)| = 1$. We conclude that reciprocal broadband splitters or couplers that yield real symmetric $S_G(s)$ cannot be used to achieve equality in (15) when $M < N$ and the loads are decoupled. We see an example of this in the next section.

## V. One Source and Multiple Loads

We probe the special case of one source more deeply with some examples in this section. We analyze constrained networks in Section V-A. Decoupled and coupled loads are examined in Sections V-B and V-C, leading to a class of non-reciprocal networks called "determinant" networks.

### A. Constrained matching network

Let a portion of the network be constrained or prescribed to have a particular structure. For example, the constrained portion might include ideal circulators or power splitters. The prescribed portion and loads then together constitute an "equivalent load" and the achievable bandwidth is determined by the characteristics of this load.

Figure 4 illustrates an example of such a structure for $M = 1$, where a prescribed $(N+1)$-port network is combined with the loads and forms an equivalent one-port load. Let the S-matrix of the prescribed network be

$$
S_0(s) = \begin{matrix} 1 \\ N \end{matrix} \begin{pmatrix} S_{0,11}(s) & S_{0,12}(s) \\ S_{0,21}(s) & S_{0,22}(s) \end{pmatrix}.
$$

Then the S-parameter of the equivalent load is

$$
\begin{aligned}
S_{L,\text{eq}}(s) = {}& S_{0,11}(s) \\
& + S_{0,12}(s)S_L(s)(I - S_{0,22}(s)S_L(s))^{-1}S_{0,21}(s). 
\end{aligned} \tag{22}
$$

The remaining unspecified portion of the network $S_1(s)$ is connected to the input port of $S_{L,\text{eq}}(s)$, and we may ask what bandwidth is attainable by $S_1(s)$.

Generally, we cannot apply the bounds directly to $S_{L,\text{eq}}(s)$. Unlike $S_L(s)$ where power is either dissipated or reflected by the loads, power in $S_{L,\text{eq}}(s)$ can also be absorbed by the constrained portion of the network when $S_0(s)$ is lossy. Let $\eta(\omega)$ denote the ratio between the power dissipated by $S_L(s)$ and the power delivered to $S_{L,\text{eq}}(s)$, defined as

$$\eta(\omega) = \frac{\|\vec{a}_2(j\omega)\|^2 - \|\vec{b}_2(j\omega)\|^2}{\|a_0(j\omega)\|^2 - \|b_0(j\omega)\|^2},$$

where $(\vec{a}_2(j\omega), \vec{b}_2(j\omega))$ and $(a_0(j\omega), b_0(j\omega))$ are the (incident, reflected) signals from $S_L(s)$ and $S_{L,\text{eq}}(s)$, respectively. Then $0 \le \eta(\omega) \le 1$ and

$$\eta(\omega) = \frac{S_{0,21}^H (I - S_{0,22}S_L)^{-H}(I - S_L^H S_L)(I - S_{0,22}S_L)^{-1}S_{0,21}}{1 - S_{L,\text{eq}}^H S_{L,\text{eq}}}, \tag{23}$$

which depends only on the prescribed network and the loads.

We assume $S_{L,\text{eq}}(s)$ satisfies (6) for some $s_0$; this does not require $S_L(s)$ to satisfy (6) for the same $s_0$. Then (10) becomes the following bound.

*Theorem 4:* For loads matched by a constrained network,

$$\int_0^\infty f(\omega) \log \sqrt{\frac{\eta(\omega)}{r^2(\omega) + \eta(\omega) - 1}} d\omega \le B_{\text{eq}} \tag{24}$$

where $B_{\text{eq}}$ is the right-hand side of (10) applied to $S_{L,\text{eq}}(s)$.

*Proof:* Using Figure 4, we define

$$r_{\text{eq}}^2(\omega) = 1 - \frac{\|a_0(j\omega)\|^2 - \|b_0(j\omega)\|^2}{\|a_1(j\omega)\|^2}$$

as the ratio between power not delivered to $S_{L,\text{eq}}(s)$ and the incident power from the source. Then it follows that

$$r^2(\omega) = 1 - \frac{\|\vec{a}_2(j\omega)\|^2 - \|\vec{b}_2(j\omega)\|^2}{\|a_1(j\omega)\|^2}$$

$$= 1 - \frac{\|a_0(j\omega)\|^2 - \|b_0(j\omega)\|^2}{\|a_1(j\omega)\|^2} \cdot \frac{\|\vec{a}_2(j\omega)\|^2 - \|\vec{b}_2(j\omega)\|^2}{\|a_0(j\omega)\|^2 - \|b_0(j\omega)\|^2}$$

$$= 1 - (1 - r_{\text{eq}}^2(\omega)) \cdot \eta(\omega).$$

Equivalently, we have $r_{\text{eq}}(\omega) = \sqrt{\frac{r^2(\omega) + \eta(\omega) - 1}{\eta(\omega)}}$.

We apply (10) to the equivalent load $S_{L,\text{eq}}(s)$. The result is an inequality on $r_{\text{eq}}(\omega)$:

$$\int_0^\infty f(\omega) \log \frac{1}{r_{\text{eq}}(\omega)} d\omega \le B_{\text{eq}},$$

where $B_{\text{eq}}$ is the right-hand side of (10) applied to $S_{L,\text{eq}}(s)$. Replacing $r_{\text{eq}}(\omega)$ with $\sqrt{\frac{r^2(\omega) + \eta(\omega) - 1}{\eta(\omega)}}$ gives us (24). ∎

We use beamforming as an example application. Let the desired amplitude and phase relationship of the antennas be denoted by the $N \times 1$ unit real-rational vector $\vec{v}(s)$ for $s = j\omega$. Then

$$S_0(s) = \begin{bmatrix} 0 & \vec{v}^T(s) \\ \vec{v}(s) & 0 \end{bmatrix} \tag{25}$$

denotes the S-matrix of a reciprocal one-to-$N$ power divider that constrains the incident signal $\vec{a}_2(j\omega)$ to be aligned with $\vec{v}(j\omega)$. We are not concerned with the so-called gain-bandwidth product of a phased array [30] which measures its ability to maintain far-field gain across a range of frequencies.

*Theorem 5:* Let the beamforming vector $\vec{v}(s)$ be a real constant unit eigenvector of $S_L(s)$. Then

$$\int_0^\infty f(\omega) \log \frac{1}{r(\omega)} d\omega \le B_{\text{eq}}, \tag{26}$$

where $B_{\text{eq}}$ is calculated using $S_{L,\text{eq}}(s) = \vec{v}^T(s)S_L(s)\vec{v}(s)$.

*Proof:* We substitute (25) into (22) and (23), where $S_{0,11} = 0$, $S_{0,12} = \vec{v}^T(s)$, $S_{0,21} = \vec{v}(s)$ and $S_{0,22} = 0$. Then (22) yields

$$S_{L,\text{eq}}(s) = \vec{v}^T(s)S_L(s)\vec{v}(s),$$

and (23) yields

$$\eta(\omega) = \frac{\vec{v}^H(I - S_L^H S_L)\vec{v}}{1 - S_{L,\text{eq}}^H S_{L,\text{eq}}} = \frac{\vec{v}^H\vec{v} - \vec{v}^H S_L^H S_L\vec{v}}{1 - \vec{v}^H S_L^H \vec{v}^* \vec{v}^T S_L\vec{v}}.$$

Because $\vec{v}(s)$ is a real unit eigenvector of $S_L(s)$, it is readily verified that $\eta(\omega) = 1$. The result then follows by applying Theorem 4. ∎

Section III-B in Part II of this paper provides an example where $B_{\text{eq}}$ in (26) varies with the choice of $\vec{v}(s)$.

We note that $\eta(\omega) = 1$ is obtained for any lossless $S_0(s)$, in which case (26) also applies. When $\eta(\omega) = 1$ and $S_L(s)$ satisfies (6) for some $s_0$, then $S_{L,\text{eq}}(s)$ also satisfies (6) for the same $s_0$; the converse is not true.

Compared with (10), (26) is smaller for $N > 1$. To see this more explicitly, let the loads be decoupled as in Section IV-A, whence $S_L(s) = S_l(s)I$. Then any real constant unit vector $\vec{v}(s)$ satisfies $\eta(\omega) = 1$, and the equivalent load satisfies $S_{L,\text{eq}}(s) = S_l(s)$. Then Theorem 5 yields

$$\int_0^\infty f(\omega) \log \frac{1}{r(\omega)} d\omega \le B_l, \tag{27}$$

where $B_l$ is the right-hand side of (10) applied to $S_l(s)$ for $M = 1$. Clearly, $B_l < NB_l$, which is the bound obtained in (15) for $M = 1$. We conclude that linear-in-$N$ behavior is not generally achieved for the beamforming structure (25). This conclusion is consistent with Theorem 3.

### B. Decoupled loads

Section IV-C says that full bandwidth cannot be achieved when reciprocal broadband splitters are used to drive $N > 1$ loads when $M = 1$. We instead consider a matching network consisting of non-reciprocal couplers in Figure 5(a) where an $(N + 1)$-port network comprising $N - 1$ circulators as the constrained part of the network is displayed. The resulting $(N + 1) \times (N + 1)$ S-matrix is

$$S_0(s) = \begin{bmatrix} 0 & 0 & \cdots & 0 & 1 \\ 1 & 0 & \cdots & 0 & 0 \\ 0 & 1 & 0 & \cdots & 0 \\ \vdots & 0 & \ddots & \ddots & \vdots \\ 0 & \cdots & 0 & 1 & 0 \end{bmatrix}.$$

The lower right $N \times N$ block of this matrix, which corresponds to $S_G(s)$, has $N$ zero eigenvalues and therefore does not satisfy the conditions of Theorem 3. Let $S_{L,\text{eq}}(s)$ be the S-matrix of the equivalent one-port load seen at the input of the circulators. Because the circulators are lossless, $\eta(\omega)=1$ in this constrained network. It is readily calculated that $S_{L,\text{eq}}(s)=[S_l(s)]^N$, which has poles and zeros at the same locations as $S_l(s)$, each with multiplicity $N$. Therefore, according to (10), the bandwidth achievable with this equivalent $S_{L,\text{eq}}(s)$ is $N$ times that achievable by $S_l(s)$. Hence, linear-in-$N$ network behavior for $N$ decoupled loads can be achieved by ideal circulators.

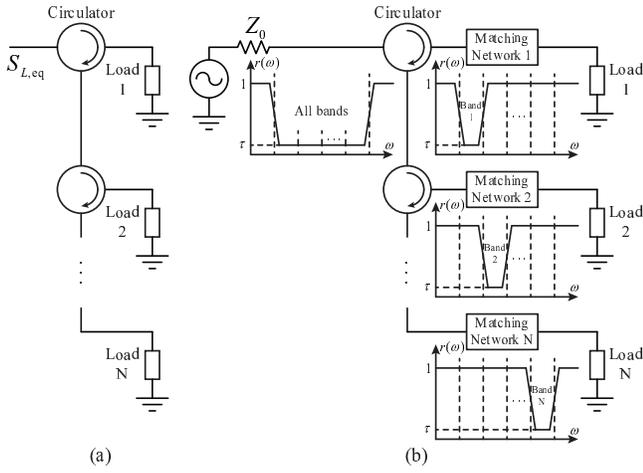

Fig. 5. (a) $N-1$ circulators are used to match $N$ decoupled loads to $M=1$ source, and achieve linear-in-$N$ behavior. We use $S_{L,\text{eq}}(s)$ to denote the S-parameter of the equivalent one-port load seen at the input of the circulators. (b) $N$ decoupled loads are driven by $M=1$ source through $N$ two-port networks and $N-1$ broadband circulators. Each of the two-ports matches $1/N$ of the total bandwidth; network $i$ passes band $i$ and reflects the remaining portions. The circulators combine all $N$ passbands and achieve a total bandwidth $N$ times of the bandwidth of each two-port network.

This circulator-based network architecture presents "multiple opportunities" for the energy that is reflected from any one load to be forwarded to the next for another attempt at transmission. We are ignoring the insertion losses associated with cascading circulators in such an arrangement.

Figure 5(b) has a structure similar to Figure 5(a) but matches $1/N$ of the total bandwidth in an orderly fashion. The first network passes the lowest portion of the band and reflects all the remaining portions, which are passed to the next circulator which matches the next portion, and so on. This structure resembles a channelizer [31] in the sense that there are multiple output ports where each port contains a portion of the total bandwidth. Although we have drawn these networks as having near-ideal flat frequency responses, this aspect is not crucial. An explicit example of this type of network for a pair of dipoles is presented in Section III-A in Part II.

### C. Circulantly-coupled loads: Determinant networks

The previous section gave a circulator-based architecture for achieving linear-in-$N$ bandwidth for $N$ uncoupled loads. We now show how to handle circulantly coupled loads such as examined in Section IV-B, by demonstrating a network architecture that converts the multiport system with S-matrix $S_L(s)$ into a single-port system with S-parameter $S_{L,\text{eq}}(s)=\det S_L(s)$. Since $\det S_L(s)$ has the same poles and zeros as $S_L(s)$ if $S_L(s)$ has no cancelling poles and zeros, $S_{L,\text{eq}}(s)$ has the same bound as $S_L(s)$. We denote any network that converts $S_L(s)$ into $\det S_L(s)$ a "determinant network". Determinant networks are linear-in-$N$.

Let $W$ be defined as in (16) and have columns $\vec{w}_1,\ldots\vec{w}_N$. Define $W_1=[\vec{w}_2\ \vec{w}_3\ \cdots\ \vec{w}_N\ \vec{0}]$ and the $(N+1)\times(N+1)$ matrix

$$S_0(s)=\begin{bmatrix} 0 & \vec{w}_N^H \\ \vec{w}_1 & W_1 W^H \end{bmatrix}. \tag{28}$$

Notice that $W_1$ is missing the column $\vec{w}_1$. Then $S_0(s)$ is constant and lossless. The following theorem says that it is also a determinant network.

*Theorem 6:* Let the network described by (28) have its $N$ outputs connected to any circulantly-coupled set of loads with S-matrix $S_L(s)$. Then its input has equivalent S-parameter $S_{L,\text{eq}}(s)=\det S_L(s)$.

*Proof:* We let $L=W^H W_1$ and use the eigenvalue decomposition $S_L(s)=W\Lambda(s)W^H$ where $\Lambda(s)=\text{diag}(\lambda_1(s),\ldots,\lambda_N(s))$ is a diagonal matrix of eigenvalues. Then (22) yields

$$\begin{aligned} S_{L,\text{eq}}(s) &= \vec{w}_N^H S_L(s)(I-W_1 W^H S_L(s))^{-1}\vec{w}_1 \\ &= \vec{w}_N^H W\Lambda(s)W^H(I-W\Lambda(s)W^H)^{-1}\vec{w}_1 \\ &= \begin{bmatrix} 0 & \cdots & 0 & \lambda_N(s) \end{bmatrix}(I-L\Lambda(s))^{-1} \\ &\quad \times \begin{bmatrix} 1 & 0 & \cdots & 0 \end{bmatrix}^T, \end{aligned}$$

where

$$I-L\Lambda(s)=\begin{bmatrix} 1 & 0 & \cdots & & 0 \\ -\lambda_1(s) & 1 & 0 & \cdots & 0 \\ 0 & -\lambda_2(s) & 1 & & 0 \\ \vdots & & \ddots & \ddots & \vdots \\ 0 & \cdots & 0 & -\lambda_{N-1}(s) & 1 \end{bmatrix}.$$

It follows that $S_{L,\text{eq}}(s)$ is $\lambda_N(s)$ times the $(N,1)$ entry of $(I-L\Lambda(s))^{-1}$. This gives $S_{L,\text{eq}}(s)=\prod_{n=1}^N \lambda_n(s)$. ∎

An intuitive explanation of the operation of the network is as follows. The network (28) orients the power from the source along the first eigenvector $\vec{w}_1$. Energy is then reflected by the loads with amplitude $\lambda_1(s)$ along $\vec{w}_1$, at which point the determinant network reflects it entirely back to the loads, but with orientation $\vec{w}_2$. This is reflected by the loads with amplitude $\lambda_2(s)$ which is returned by the network to the loads reoriented along $\vec{w}_3$, and so on. The last eigenvector $\vec{w}_N$ reflected by the loads is returned to the source. The result is an overall S-parameter with value $\prod_{n=1}^N \lambda_n(s)$, which is the determinant of $S_L(s)$. This description also explains why a determinant network is not unique.

For example, for $N=2$,

$$S_0(s)=\begin{bmatrix} 0 & \frac{1}{\sqrt{2}} & -\frac{1}{\sqrt{2}} \\ \frac{1}{\sqrt{2}} & \frac{1}{2} & \frac{1}{2} \\ \frac{1}{\sqrt{2}} & -\frac{1}{2} & -\frac{1}{2} \end{bmatrix}. \tag{29}$$

As can by seen by this S-matrix, the network works by first exciting the even mode of the coupled loads. Any reflected power, which is also in an even mode, is then converted into an odd-mode excitation.

One possible realization of (29) is given in Figure 6, which utilizes an ideal transformer and gyrator. A gyrator with resistance $Z_0$ is transparent to forward propagating waves but behaves as a $\pi$ phase shift for reverse waves. The network in Figure 6 converts the wave incident on the transformer primary into an even-mode excitation for the loads on its secondaries. Any power reflected by the loads, which is also in an even mode, is converted into an odd mode by the gyrator. The secondaries of the transformer acts like an open circuit to this odd-mode signal. The signal is therefore reflected by the transformer back to the loads as an odd-mode excitation. Finally, any odd-mode excitation reflected by the antennas is then converted by the gyrator into an even-mode signal and passed by the transformer back to the source.

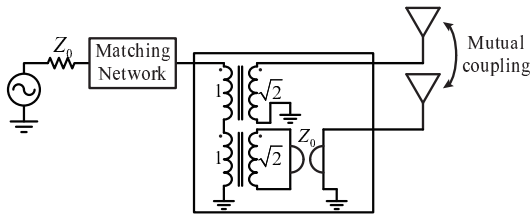

Fig. 6. Realization of the network (29) that includes a transformer and non-reciprocal gyrator connected to two coupled antennas.

The network in Figure 6 can achieve the bound in (10) for a single source and two identical loads that are reciprocally coupled. This network would also work well for the decoupled loads in Section IV-A since decoupled loads are trivially reciprocal. Note that the circulator structure proposed in Figure 5 is a determinant network when the loads are decoupled but is otherwise not. Hence, the network in Figure 6 generalizes the networks in Figure 5 for $N = 2$.

## VI. Conclusions

We have presented bandwidth analyses for multiport RF systems using bandwidth upper bounds that apply to an arbitrary number of sources and loads, and allows arbitrary coupling between the loads. Conditions for achieving equality in the bounds were discussed. We demonstrated that the bandwidth bounds scale generally as $N/M$ for $M$ sources and $N$ loads. We focused on loads whose scattering matrices can be expressed analytically with rational functions. The case of one source and many loads was examined deeply.

In Part II of this paper, we apply the upper bounds to realistic loads whose scattering parameters are expressed numerically and therefore need to be approximated by rational functions. The accuracy needed of such approximations is analyzed. Some of the effects of coupling are examined in detail, and complete derivations of the bounds are also provided.

## Appendix
### Proof of Theorem 3

Let $\vec{v}(s)$ be the eigenvector of $S_L(s)$ and $S_G(s)$ associated with eigenvalues $\lambda_L(s)$ and $\lambda_G(s)$. Because $S_L(s), S_G(s)$ are bounded matrices, $\lambda_L(s), \lambda_G(s)$ are bounded functions. Since $|\lambda_G(j\omega)| = 1$, $\lambda_G(s)$ is an all-pass function with all poles in the LHP and zeros in the RHP. Moreover, because $I - S_L^T(-s)S_L(s)$ is full normal rank and $S_L(s)$ satisfies

(6), $\lambda_L(s)$ is non-constant and satisfies $\lambda_L(-s_0)\lambda_L(s_0) = 1$ where $s_0$ is defined in (6). Thus, $|\lambda_L(s)| < 1$ for $\text{Re}\{s\} > 0$, $|\lambda_G(s)| \geq 1$ for $\text{Re}\{s\} < 0$ and $|\lambda_G(s)| \leq 1$ for $\text{Re}\{s\} > 0$; the equalities holds if $\lambda_G(s)$ is a constant.

We start by showing $\vec{v}^H(j\omega)S_G(j\omega) = \lambda_G(j\omega)\vec{v}^H(j\omega)$. Since $\lambda_G(j\omega)$ and $\vec{v}(j\omega)$ are an eigenvalue-eigenvector pair of $S_G(j\omega)$,

$$\vec{v}^H(j\omega)(S_G(j\omega) - \lambda_G(j\omega)I)\vec{v}(j\omega) = 0.$$

Suppose $\vec{u}^H(j\omega) = \vec{v}^H(j\omega)(S_G(j\omega) - \lambda_G(j\omega)I) \neq 0$, then $\vec{u}(j\omega) \perp \vec{v}(j\omega)$, and

$$\|\vec{v}^H(j\omega)S_G(j\omega)\| = \|\lambda_G(j\omega)\vec{v}^H(j\omega) + \vec{u}^H(j\omega)\|$$
$$> \|\lambda_G(j\omega)\vec{v}^H(j\omega)\| = \|\vec{v}^H(j\omega)\|.$$

But because $S_G(s)$ is bounded, all singular values of $S_G(j\omega)$ are no greater than one. This contradicts the inequality above. Hence, $\vec{u}^H(j\omega) = 0$, and we have $\vec{v}^H(j\omega)S_G(j\omega) = \lambda_G(j\omega)\vec{v}^H(j\omega)$.

Let $\vec{v}'(s)$ be the extension of $\vec{v}^*(j\omega)$ to the WCP. We then show that Condition 4 cannot be achieved when Condition 3 is satisfied. We need to show $\vec{v}'^T(s)(S_L^T(-s) - S_G(s)) = (\lambda_L(-s) - \lambda_G(s))\vec{v}'^T(s) = 0$ for some $s$ in the RHP. It suffices to show $\lambda_L(-s) - \lambda_G(s) = 0$.

Since $|\lambda_L(-s)| < 1$ and $|\lambda_G(s)| \geq 1$ for $\text{Re}\{s\} < 0$, all zeros of $\lambda_L(-s) - \lambda_G(s) = 0$ must locate in $\text{Re}\{s\} \geq 0$. Both $\lambda_L(s)$ and $\lambda_G(s)$ are rational functions, so we let their degrees be $n_1$ and $n_2$, respectively. Since all the poles of $\lambda_L(s)$ and $\lambda_G(s)$ are in the LHP, the poles of $\lambda_L(-s)$ and $\lambda_G(s)$ do not coincide. Hence $\lambda_L(-s) - \lambda_G(s)$ has degree $n_1 + n_2$. Moreover, $|\lambda_L(j\omega)| = 1$ has at most $n_1$ solutions; this is because $1 - \lambda_L(-s)\lambda_L(s)$ has degree at most $2n_1$, and any zeros of $1 - \lambda_L(-s)\lambda_L(s)$ on the imaginary axis have multiplicity at least two. Since $|\lambda_G(j\omega)| = 1$ for all $\omega$, $\lambda_L(-s) - \lambda_G(s)$ has at most $n_1$ zeros on the imaginary axis. This leaves at least $n_2$ zeros in the RHP, and we have proven the theorem when $n_2 \geq 1$.

When $n_2 = 0$, we prove by contradiction. Suppose all $n_1$ zeros of $\lambda_L(-s) - \lambda_G(s)$ are on the imaginary axis. Then the zeros of $1 - \lambda_L(-s)\lambda_L(s)$ are at the same locations as the zeros of $\lambda_L(-s) - \lambda_G(s)$. Because we assume $\lambda_L(-s_0)\lambda_L(s_0) = 1$ for some $s_0$, $s_0$ must be one of the zeros of $\lambda_L(-s) - \lambda_G(s)$. Thus, we have $\lambda_G(s_0) = \lambda_L(-s_0)$, and $1 - \lambda_G(s_0)\lambda_L(s_0) = 0$. This contradicts Condition 3. Hence, $\lambda_L(-s) - \lambda_G(s)$ has zeros in the RHP when $n_2 = 0$. Thus, Condition 4 is not met, and since this condition is necessary for achieving the bounds, the bounds cannot be met with equality. ∎

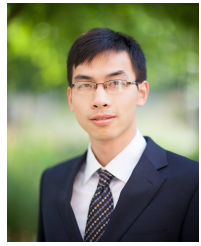

**Ding Nie** Ding Nie (M'16-) was born in Nanchang, Jiangxi, Peoples Republic of China. He received the B.S. degree in electronic engineering from Shanghai Jiao Tong University, Shanghai, in 2010. He received the M.S. degree and the Ph.D. degree in electrical engineering from the University of Notre Dame, Notre Dame, IN, in 2016. He then joined Apple Inc., Cupertino, CA, USA, in 2016. He won the 2016 Outstanding Young Author Award for the paper he co-authored in the IEEE Transactions on Circuits and Systems. His research interests include communications, radio-frequency circuits and antennas.

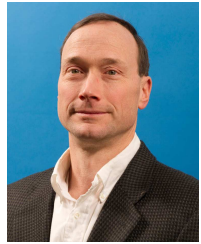

**Bertrand M. Hochwald** Bertrand M. Hochwald (F'08-) was born in New York, NY. He received his undergraduate education from Swarthmore College, Swarthmore, PA. He received the M.S. degree in electrical engineering from Duke University, Durham, NC, and the M.A. degree in statistics and the Ph.D. degree in electrical engineering from Yale University, New Haven, CT.

From 1986 to 1989, he worked for the Department of Defense, Fort Meade, MD. After completing graduate school he was a Research Associate and Visiting Assistant Professor at the Coordinated Science Laboratory, University of Illinois, Urbana-Champaign. In September 1996, he joined the Mathematics of Communications Research Department at Bell Laboratories, Lucent Technologies, Murray Hill, NJ where he was a Distinguished Member of the Technical Staff.

In 2005 he joined Beceem Communications, Santa Clara, CA, as their Chief Scientist, and in 2009 as Vice-President of Systems Engineering. He served concurrently as a Consulting Professor in Electrical Engineering at Stanford University, Palo Alto, CA. In 2011, he joined the faculty at the University of Notre Dame as Freimann Professor of Electrical Engineering.

He received several achievement awards while employed at the Department of Defense and the Prize Teaching Fellowship at Yale University. He has served as an Editor for several IEEE journals and given plenary and invited talks on various aspects of signal processing and communications. He has co-invented several well-known multiple-antenna techniques, including a differential method, linear dispersion codes, and multi-user vector perturbation methods. He has forty-five patents.

He received the 2006 Stephen O. Rice Prize for the best paper published in the IEEE Transactions on Communications. He co-authored a paper that won the 2016 Best Paper Award by a young author in the IEEE Transactions on Circuits and Systems. He is listed as a Thomson Reuters Most Influential Scientific Mind in 2014 and 2015. He is a Fellow of the IEEE.

# Bandwidth Analysis of Multiport Radio-Frequency Systems—Part II

Ding Nie, *Member, IEEE,* and Bertrand M. Hochwald, *Fellow, IEEE*

*Abstract*—We analyze the bandwidth over which power can be transferred from multiple radio-frequency sources to multiple loads through a passive multiport matching network. This is the second part of a two-part paper. In the first part we introduce broadband multiport matching upper bounds and apply them to determine the bandwidth of loads and network structures whose scattering parameters can be expressed analytically with rational functions. In this second part, we apply the bounds to loads, such as antennas, whose scattering parameters are obtained by measurement or simulation. We focus on the effects of coupling on bandwidth. Since the bounds require frequency responses that are rational functions, we provide guidelines on how to obtain rational approximations for arbitrary loads. Complete derivations of the bounds are also provided.

*Index Terms*—Bandwidth, Bode-Fano bounds, broadband matching bounds, non-reciprocal networks, passive matching networks, radio-frequency coupling

## I. INTRODUCTION

We analyze the bandwidth over which power can be transferred from $M$ sources through a passive matching network to $N$ loads in radio-frequency (RF) systems. Our principal tools include broadband matching upper bounds, which are presented in Part I of this paper [1], and briefly repeated herein for easy reference. These bounds extend the classical Bode-Fano results [2], [3], which apply to a single source and load. The bounds depend only on the scattering matrix (S-matrix) of the loads. The loads may be electromagnetically coupled to each other.

In this part, we apply the bounds to loads whose S-matrix is expressed numerically, found through either measurements or simulations. The bounds require rational models of the loads that can be obtained through fitting of the numerical data. We provide guidelines for assessing the accuracy of a rational model.

Of particular interest is the effect of coupling on the bandwidth of multiple antennas. One example demonstrates that the bandwidth bound of a pair of dipoles fluctuates non-monotonically with the distance between them, and, with the right amount of coupling, can be significantly greater than that of decoupled dipoles. We focus on the "input bandwidth" of the loads, and are not concerned with how efficiently the loads use their input power.

D. Nie is with Apple Inc., Cupertino, CA 95014, USA (e-mail: kirknie@gmail.com).

B. M. Hochwald is with the Department of Electrical Engineering, University of Notre Dame, Notre Dame, IN 46556, USA (e-mail: bhochwald@nd.edu).

This work was supported, in part, by NSF grants CCF-1403458 and ECCS-1509188.

We also include complete derivations of the bounds. These derivations are not needed to apply the bounds but are of use in understanding how the conditions for equality in the bound apply.

The notations and definitions are identical to those in Part I. We assume $M$ decoupled sources with characteristic impedance $Z_0$ drive $N$ loads that have a frequency-dependent $N \times N$ S-matrix $S_L(s)$, where $s = \sigma + j\omega$ is the extension of frequency $j\omega$ to the whole complex plane (WCP). An $(M + N)$-port passive matching network is inserted between the sources and loads. Our measure of the quality of matching is the power loss ratio $r^2(\omega) \in [0, 1]$, defined as the ratio between the total power lost, including insertion and return losses, and the total incident power from the sources at frequency $\omega$. The insertion loss is the power dissipated in the matching network; the return loss is the power reflected to the sources. Since the loads are potentially coupled, the quantity $r^2(\omega)$ captures the possibility that incident power from one source can reflect back to itself or to other sources. We then define the bandwidth $\omega_{\mathrm{BW}}(\tau, \omega_d)$ as the frequency range for which $r(\omega)$ (the positive square root of $r^2(\omega)$) is no greater than a threshold $\tau > 0$ in the vicinity of a design frequency $\omega_d$. The mathematical definitions of $r(\omega)$ and $\omega_{\mathrm{BW}}(\tau, \omega_d)$ can be found in Part I.

We generally employ the properties of S-matrices of passive real networks to prove the bounds; these properties include real-rational, Hurwitzian and bounded. We also use the definitions of poles and zeros of rational matrices. We use LHP to denote the left-half complex plane ($\mathrm{Re}\{s\} < 0$) and RHP to denote the right-half ($\mathrm{Re}\{s\} > 0$).

### A. Summary of bounds

There are three bounds which are summarized as follows. If

$$S_L^T(-s_0)S_L(s_0) = I \tag{1}$$

is satisfied for some $\mathrm{Re}\{s_0\} \geq 0$, then the following inequality holds for any passive matching network:

$$\int_0^\infty f(\omega) \log \frac{1}{r(\omega)} d\omega \leq B, \tag{2}$$

where $f(\omega)$ and $B$ depend on the location of $s_0$ and are given in Table I. To compute the bound $B$, the poles and zeros of $S_L(s)$, denoted as $p_{L,i}, z_{L,i}$, are also needed. We refer to rows 1, 2, and 3 in Table I as Bounds 1, 2, and 3.

The bounds depend on the poles and zeros of the real-rational matrix $S_L(s)$, which we explain how to obtain in

| | |
|---|---|
| $s_0 = j\omega$ | $f(\omega) = \frac{1}{2}[(\omega_0 - \omega)^{-2} + (\omega_0 + \omega)^{-2}]$ |
| | $B = \frac{-\pi}{2M}\left[\sum_i (p_{L,i} - j\omega_0)^{-1} + \sum_i (z_{L,i} + j\omega_0)^{-1}\right]$ |
| $\mathrm{Re}\{s_0\} > 0$ | $f(\omega) = \frac{1}{2}\mathrm{Re}[(s_0 - j\omega)^{-1} + (s_0 + j\omega)^{-1}]$ |
| | $B = \frac{-\pi}{2M}\log\left|\det S_L(s_0) \cdot \frac{\prod_i(s_0 + z_{L,i})}{\prod_i(s_0 - z_{L,i})}\right|$ |
| $s_0 = \infty$ | $f(\omega) = 1$ |
| | $B = \frac{-\pi}{2M}\left(\sum_i p_{L,i} + \sum_i z_{L,i}\right)$ |

Section II. We focus on the steps needed to fit realistic loads with rational models. Bandwidth analyses of various loads are then presented in Section III. The bounds are corollaries of Theorems 2 and 3, which are presented in Section IV. We conclude in Section V.

## II. OBTAINING $S_L(s)$ AND ITS POLES AND ZEROS

The matrix $S_L(s) = (Z_L(s) + Z_0 I)^{-1}(Z_L(s) - Z_0 I)$ is uniquely determined if the impedance matrix of the loads $Z_L(s)$ is known analytically. For many realistic loads, this is generally impossible. Instead, either $Z_L(j\omega)$ or $S_L(j\omega)$ is often obtained as the result of a numerical simulation or measurement, where $\omega$ is contained to some frequency range of interest. The next section gives an example of how to use the results of simulation to find $S_L(s)$.

### A. Finding $S_L(s)$ from numerical simulations of loads

The following steps can be used to find $S_L(s)$:

a) Measure or simulate the loads in the frequency range of interest $[\omega_1, \omega_2]$. Denote the measured response $S'_L(j\omega)$.
b) Find a passive rational $S_L(s)$ such that $S_L(j\omega)$ is "close enough" to $S'_L(j\omega)$ for $\omega \in [\omega_1, \omega_2]$.

Step a) can be done with standard modeling software such as Ansys HFSS in the case of simulations, or a network analyzer in the case of measurements. Step b) can be accomplished by fitting rational functions to the individual entries of $S'_L(j\omega)$ using, for instance, the Matrix Fitting Toolbox [4]–[8] in MATLAB. We have more to say about this fitting in the next section.

To compute the right-hand side of (2):

c) Find an $s_0$ where (1) is satisfied for the computed $S_L(s)$.
d) Calculate the poles and zeros $p_{L,i}, z_{L,i}$ of $S_L(s)$.

Step c) requires us to solve $S_L^T(-s_0)S_L(s_0) = I$. For $s_0 = j\omega_0$, we can instead solve $|\det S_L(j\omega_0)| = 1$. For Step d), $p_{L,i}, z_{L,i}$ can be obtained from the definitions of poles and zeros of matrices. In some cases the poles and zeros of $S_L(s)$ coincide with the poles and zeros of $\det S_L(s)$. One way of checking this is presented in [9], which we do not repeat here. Then $p_{L,i}, z_{L,i}$ can be obtained by applying root-finding algorithms to $1/\det S_L(s) = 0$ and $\det S_L(s) = 0$, respectively. The examples shown in Sections III–B–III–E use this approach.

Multiple rational models may exist within the error tolerance of $S'_L(j\omega)$ for $\omega \in [\omega_1, \omega_2]$. Each model could satisfy

(1) for different $s_0$, and generate different bounds. When these models yield consistent results, one can build confidence that the bounds and models are physically meaningful. When these models contradict each other, further investigation is needed to determine the source of the inconsistency. An example of multiple models is shown in Section III-C.

### B. Analysis of distributed-element loads using rational models

Since $S_L(s)$ is a real-rational matrix, accurate modeling of lumped-circuit loads is straightforward and can be done without error. However, many loads such as antennas, transmission lines, and other distributed-element systems include time-delays, stubs, and other structures that are often modeled using non-rational functions of $s$. A real-rational $S_L(s)$ is then needed that "approximates" the loads with sufficient accuracy to obtain a meaningful bandwidth bound.

The numerical and analytical approximation of distributed-element circuits by lumped circuits over a fixed bandwidth is a topic that has been studied in many contexts. An early example is [10]. A summary of some practical techniques can be found in [11]. Recent advances in antenna modeling include [12]. Other fields such as control engineering use rational fitting extensively [13].

The mathematical problem of rational approximation and modeling is addressed by Runge's theorem in complex analysis [14], which states that any analytical function on a subset of the complex plane can be fit arbitrarily precisely with rational functions. For example, the Padé approximation [15], [16] is often used to model time delays using rational functions over a frequency band of interest.

We are interested in obtaining an upper bound on

$$\int_{\omega_1}^{\omega_2} f(\omega) \log \frac{1}{r'(\omega)} d\omega, \tag{3}$$

where $r'(\omega)$ is the loss experienced when the matching network is used with the actual loads $S'_L(j\omega)$. The process outlined in Section II-A creates a rational approximation $S_L(j\omega)$ for $\omega \in [\omega_1, \omega_2]$. For an arbitrary matching network, (2) gives

$$\int_{\omega_1}^{\omega_2} f(\omega) \log \frac{1}{r(\omega)} d\omega \le \int_0^\infty f(\omega) \log \frac{1}{r(\omega)} d\omega \le B, \tag{4}$$

where $B$ is computed from $S_L(s)$ and $r(\omega)$ is the power loss ratio experienced with $S_L(s)$. The difference between the left-hand sides of (3) and (4) is

$$\int_{\omega_1}^{\omega_2} f(\omega) \log \frac{r(\omega)}{r'(\omega)} d\omega,$$

and represents the error in the integral introduced by the approximation of $S'_L(j\omega)$. Then choosing $S_L(s)$ such that

$$\int_{\omega_1}^{\omega_2} f(\omega) \log \frac{r(\omega)}{r'(\omega)} d\omega \le \delta B \tag{5}$$

for some desired tolerance $\delta B > 0$ yields the bound

$$\int_{\omega_1}^{\omega_2} f(\omega) \log \frac{1}{r'(\omega)} d\omega \le B + \delta B. \tag{6}$$

The following theorem provides a first-order approximation of $\delta B$ for $S_L(s)$ that is close to $S'_L(j\omega)$ in the frequency band of interest.

*Theorem 1:* Let $\delta S_L(j\omega) = S_L(j\omega) - S'_L(j\omega)$ and define

$$\rho(\omega) = \frac{2\sigma_{\delta,\max}(\omega)}{1 - \sigma'^2_{L,\max}(\omega)} \cdot \sqrt{1 + \frac{\sigma'^2_{L,\max}(\omega) - \sigma'^2_{L,\min}(\omega)}{(1 - \sigma'_{L,\max}(\omega))^2}}, \quad (7)$$

where $\sigma_{\delta,\max}(\omega)$ is the maximum singular value of $\delta S_L(j\omega)$, and $0 \leq \sigma'_{L,\min}(\omega) < \sigma'_{L,\max}(\omega) < 1$ are the minimum and the maximum singular values of $S'_L(j\omega)$, respectively. Then

$$\delta B \to \int_{\omega_1}^{\omega_2} \frac{f(\omega)}{2} \log\left(1 + \frac{1 - r'^2(\omega)}{r'^2(\omega)} \cdot \rho(\omega)\right) d\omega \quad (8)$$

as $\|\delta S_L(j\omega)\|_F \to 0$ for $\omega \in [\omega_1, \omega_2]$.

*Proof:* See Appendix A. ∎

In (8), $f(\omega)$ is given in Table I, and $\rho(\omega)$ is determined by $S'_L(j\omega)$ and $\delta S_L(j\omega)$. The notation $\|\cdot\|_F$ refers to the Frobenius norm (sum of squared magnitude of entries) of a matrix. In the limit of a perfect model when $\delta S_L(j\omega) = 0$, then $\rho(\omega) = 0$ and $\delta B = 0$. In general, $\delta B$ depends on $r'(\omega)$, which depends on the matching network. To eliminate the dependence of $\delta B$ on $r'(\omega)$, we choose the desired goal of $r'(\omega) \approx \tau$ for $\omega \in [\omega_1, \omega_2]$, whence

$$\delta B \approx \int_{\omega_1}^{\omega_2} \frac{f(\omega)}{2} \log\left(1 + \frac{1 - \tau^2}{\tau^2} \cdot \rho(\omega).\right) d\omega. \quad (9)$$

Note that even if $r'(\omega)$ does not achieve $\tau$ for $\omega \in [\omega_1, \omega_2]$, (9) is still a useful quantity because it upper-bounds $\delta B$.

The steps for evaluating goodness-of-fit are then:

e) Compute $\delta B$ in (9) numerically.

f) Evaluate if the desired tolerance on $\delta B/B$ is met.

If the desired tolerance is not met, a smaller $\delta S_L(j\omega)$ is needed. Finally,

g) Compute the bound (6).

Theorem 1 can also be used to examine cases where the rational fit may fail to generate an accurate bound. For example, if $\sigma'_{L,\max}(\omega) \approx 1$ in the neighborhood of some $\omega$, meaning $S'_L(j\omega)$ is very reflective, (7) indicates that $\rho(\omega)$ becomes large. Any attempt to make $r'(\omega)$ small around this $\omega$ could then potentially make $\delta B$ in (8) large. Thus, attempting to match a load where it is naturally very reflective may lead to a loose bound.

## C. Hazard of over-fitting

In attempting to make $\delta B/B$ small, care should be taken not to "over-fit" the loads. Over-fitting or over-modeling occurs when the degrees of the polynomials in the numerator and denominator of the rational functions in $S_L(s)$ are larger than needed to obtain the desired accuracy. As the polynomial degrees are increased, the relative error $\delta B/B$ is decreased since $\delta B$ is decreased. But there is also the hazard that $B$ is made unnecessarily large because of the excessive number of poles and zeros in $S_L(s)$. As a result, (6) becomes loose and Condition 4 for achieving $B$ cannot be satisfied. There is a tension between making the rational model accurate while still having a small number of poles and zeros.

One approach to determine if the system is over-fitted is to examine the poles and zeros of the entries of $S_L(s)$. Any poles and zeros that nearly cancel can have a small effect on $\delta B$ but a large effect on $B$; they should be removed and $\delta B/B$ re-checked. Simpler models yield tighter bounds, and thus there is an incentive to find the minimal model that adequately captures the frequency response of the loads in the band of interest. A detailed discussion of modeling poles is found in [17]. A discussion of the sensitivity of poles and zeros to perturbations of the data is found in [18].

In the remainder, we use the notation $r(\omega)$ when we are considering the rational model, and we use $r'(\omega)$ when we are considering the actual loads. In Section III-D, we present an example where time delay in a circuit is modeled using rational functions. Section III-E computes (6) for four realistic coupled antennas.

## III. Bandwidth Analysis of Antennas

We now calculate broadband matching bounds for five examples, all involving amplifiers driving antennas. Aspects of coupling and rational approximation are emphasized. The first three examples assume that the rational models are accurate and therefore compute only (2), while the last two examples incorporate the effects of rational fitting and therefore compute (6). We begin with two identical decoupled dipoles whose rational models are taken from [9].

## A. Two decoupled dipoles

We design a broadband network of the type in Figure 5(b) in Part I for two decoupled dipoles ($N = 2$) with one source ($M = 1$). From Theorem 1 in Part I, the bound for two decoupled dipoles is twice as large as for a single dipole when there is only one source. The geometry of a dipole is shown in Figure 1(a), which is half-wavelength at 2.4 GHz. In [9], the same dipole is simulated in the range 1–5 GHz, and is modeled using an S-parameter model $S_l(s)$ normalized by characteristic impedance $Z_0 = 50\ \Omega$ that satisfies $S_l(\infty) = -1$. We apply the model here, and write the S-matrix of two such dipoles as $S_L(s) = \mathrm{diag}(S_l(s), S_l(s))$.

Clearly, $S_L(\infty) = -I$, and (1) is satisfied at $s_0 = \infty$. The poles and zeros of $S_l(s)$ are listed in Table III in [9], and Bound 3, using (2), for $M = 1$ gives

$$\int_{\omega_1}^{\omega_2} \log \frac{1}{r(\omega)} d\omega \leq \int_0^\infty \log \frac{1}{r(\omega)} d\omega \leq 4.93 \times 10^{10}, \quad (10)$$

where $\omega_1 = 2\pi \times 10^9$ and $\omega_2 = 10\pi \times 10^9$ are the lower and upper modeling frequencies.

We are interested in matching the decoupled dipoles to one source in the 2–4 GHz range. The structure in Figure 2 is employed, where Matching Network 1 is tuned for the 2–3 GHz band and Matching Network 2 is tuned for the 3–4 GHz band and one circulator is used. These two-port networks are designed using the Real Frequency Technique [19], followed by a realization using Darlington synthesis [20]. In Figure 1(b), the dashed and dash-dot curves show $r(\omega)$ of each network when connected by itself to a dipole (typically through a balun that is not shown). The $r(\omega)$ of the entire system, including circulator, is the product of the $r(\omega)$ achieved by each, and is shown as the solid curve. We seek a power reflection ratio of $r(\omega) \leq 0.2$ ($-14$ dB). From the figure, we see that the bandwidth achievable is $\omega_{\mathrm{BW}}(0.2, 3\,\mathrm{GHz}) = 2.35$ GHz. The achieved integral is

$$\int_{\omega_1}^{\omega_2} \log \frac{1}{r(\omega)}\, d\omega = 4.58 \times 10^{10}. \quad (11)$$

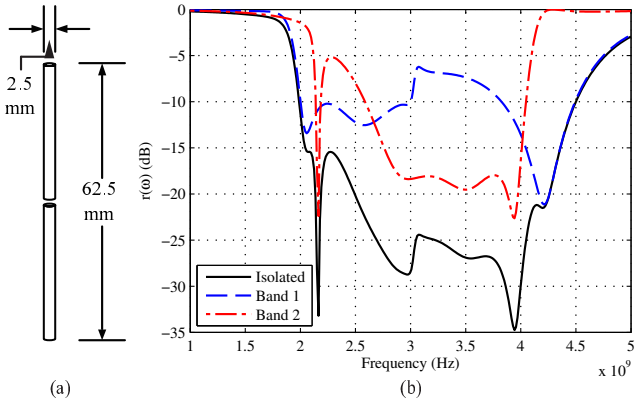

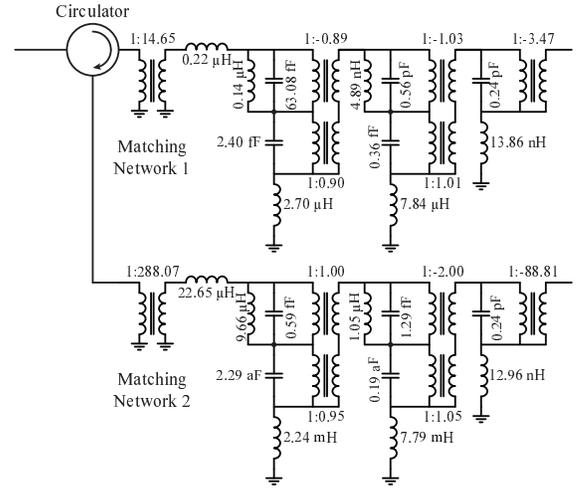

Fig. 1. (a) Geometry of a dipole that is half-wavelength at 2.4 GHz. (b) Two decoupled dipoles are matched by the structure shown in Figure 2. Matching Network 1 handles 2–3 GHz and Network 2 handles 3–4 GHz; their resulting $r(\omega)$'s are given in the blue dashed and red dash-dot curves, respectively. The overall system $r(\omega)$ is shown by the solid curve.

Fig. 2. Matching network of the type in Figure 5(b) in Part I is used to connect a source (on the left) to two decoupled dipoles (connected on the right, typically through baluns that are not shown). Matching networks 1 and 2 are designed using the Real Frequency Technique [19], followed by realizations using Darlington synthesis [20].

There is only a $3.51 \times 10^9$ gap between (10) and (11), which is due entirely to shaping loss defined in Part I, Section III-C. We therefore achieve excellent performance for one source and two dipoles with this non-reciprocal network. According to Theorem 3 in Part I, since $M < N$, using a reciprocal splitter or coupler as part of the network would not perform as well.

An experimental measurement of the effectiveness of the matching network in Figure 2 would require standard S-parameter measurements by a network analyzer at the input port of the circulator. Since there is only one source, no averaging is needed over phase differences, as described in Part I, Section II-C. By the definition of $r(\omega)$, and because the network is (theoretically) lossless, small values of $r(\omega)$ in 2–4 GHz imply that the power from the source is being accepted by the antennas.

We also note that for two sources ($M = 2$), we would be able to achieve only half the bandwidth for the same $-14$ dB threshold. Clearly, the matching network for two sources would differ considerably from the one presented in Figure 2 and could be reciprocal. We do not design such a network here.

### B. Two coupled dipoles

To illustrate the effect of coupling between two parallel dipoles, we examine their bandwidth as a function of the distance between them. We assume each dipole has the same 2.4 GHz half-wavelength structure as Figure 1(a). The center feeding points of the parallel dipoles are at the same vertical level, and the distance between them is $d$, as measured at any place along their lengths. For each value of $d$, we apply the modeling recipe for the S-matrix $S_L(j\omega)$ detailed in Section II in the range 1–5 GHz. Since we wish to compare the bounds for coupled dipoles with (10), we model their S-matrices using six poles and six zeros, and enforce $S_L(\infty) = -I$. Then the resulting model $S_L(s)$ satisfies (1) at $s_0 = \infty$ for every $d$, and Bound 3 can be applied to compare with (10).

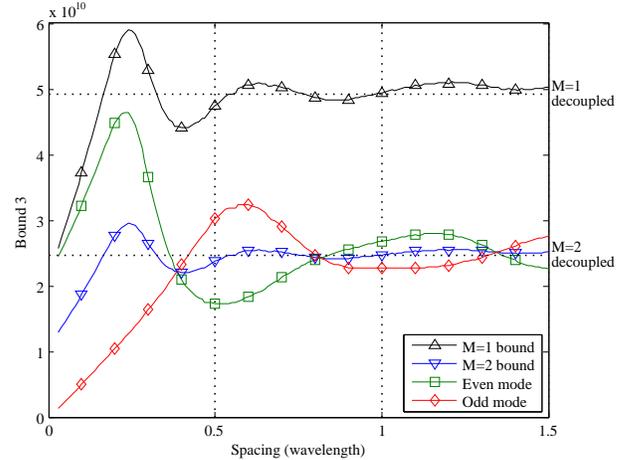

Fig. 3. Bound 3 for two parallel dipoles versus spacing $d$, for $d$ between $0.03\lambda$ and $1.5\lambda$ ($\lambda = 125$ mm is the wavelength at 2.4 GHz). Each dipole has the structure shown in Figure 1(a), the center feeding points of the dipoles are at the same vertical level, and the distance $d$ is maintained along their lengths. The horizontal dotted lines indicate Bound 3 for two decoupled dipoles. Also shown are the even- and odd-mode beamforming bounds.

We let $d$ range from $0.03\lambda$ to $1.5\lambda$, with step size $0.01\lambda$, where $\lambda = 125$ mm is the wavelength at 2.4 GHz. The poles and zeros $p_{L,i}, z_{L,i}$ are computed from $S_L(s)$, and Bound 3 is computed for each $d$. For $M = 1$ and $M = 2$, the bounds are shown in Figure 3 by the black and blue curves. The black curve values are twice the blue curve. Horizontal dotted lines indicate the bounds for decoupled dipoles.

Figure 3 shows oscillatory behavior and suggests that coupled dipoles can have large bandwidths at certain distances from each other. The maxima appear near $0.24\lambda$ and are 20% larger than their decoupled-dipole counterparts. As $d$ increases, the bounds approach the decoupled-dipole limits. On the other hand, when $d$ approaches zero, the two dipoles merge into a single dipole, and the bounds in Figure 3 approach the bounds where $M$ sources drive a single dipole.

We already know from Part I, Section V-A, that using these dipoles in a beamformer configuration with $M = 1$ generally cannot achieve the bound for any $d$. But the beamformer configuration is still worth analyzing briefly. There are two beamformers that are of special interest: even-mode and odd-mode, corresponding to $\vec{v}(s) = [1/\sqrt{2}, 1/\sqrt{2}]^T$ and $[1/\sqrt{2}, -1/\sqrt{2}]^T$ in (25) in Part I. Since the dipoles are reciprocal and have a symmetric structure, the resulting $S_L(s)$ is symmetric circulant, and both even and odd $\vec{v}(s)$ are real constant unit eigenvectors of $S_L(s)$. Thus Theorem 5 in Part I applies.

Figure 3 shows the results. Both even and odd-mode bounds are strictly smaller than the bound for $M = 1$, as expected. Their sum, however, equals the $M = 1$ bound because the incident and reflected waves to and from the loads are orthogonal for the two modes. Therefore, the modes can be treated as decoupled loads, each with its own equivalent S-parameter and bound.

### C. Multiple models for the same loads

For a set of loads with a measured $S'_L(j\omega)$, more than one model $S_L(s)$ may be created within a given error tolerance. To illustrate this, we take the parallel dipoles in Section III-B with $S'_L(j\omega)$ measured in 1–5 GHz as examples. For each $d$, we model the S-matrix using the Matrix Fitting Toolbox with six poles and six zeros and enforce $S_L(0) = I$. The resulting $S_L(s)$ then satisfies (1) at $s_0 = 0$, and Bound 1 is applied to the coupled dipoles. We contrast the results with Section III-B, where $S_L(\infty) = -I$ is enforced and Bound 3 applies.

For $M = 1$, Figure 4 compares Bound 1 (left $y$-axis) with Bound 3 (right $y$-axis) for the parallel dipoles when $d$ varies from $0.03\lambda$ to $1.5\lambda$. The horizontal dotted line indicates the limiting values as the dipoles become decoupled. Both curves show similar trends when the dipoles are closely spaced, and suggest a bandwidth peak at $0.24\lambda$. For $d > 0.32\lambda$, the curves seem out of phase with each other. Further exploration is needed to see if there is any physical significance to this.

### D. Rational approximation of time delays

Any characterization of physically separated antennas should account for the time of propagation of the signal between the antennas. Time delays associated with this distance can be modeled as a non-rational (exponential) function of $s$.

A simple idealized analytical model of time delay is shown in Figure 5, which consist of two RC loads and a coupling branch between them. The coupling is capacitive and includes a delay component. We examine our ability to match to this load in the 100–400 MHz band. As a reference value, without the time delay (zero delay), the S-matrix of the loads is real-rational and satisfies $S_L(\infty) = -I$, has two poles at $-8.00 \times 10^8$ and $-4.44 \times 10^8$, and has a zero at 0 with multiplicity two. We apply Bound 3 and obtain $B = 9.77 \times 10^8$.

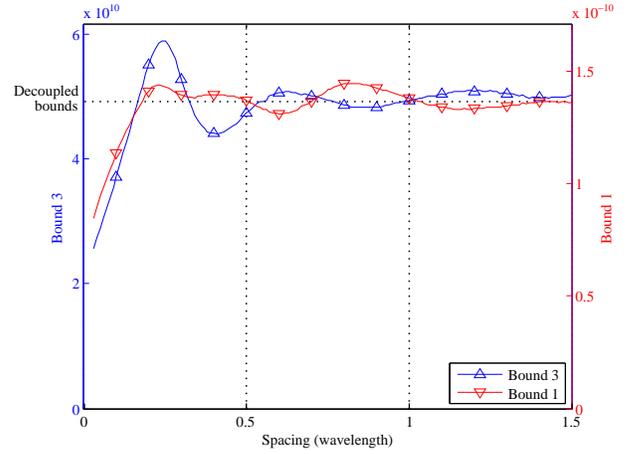

Fig. 4. For one source and two parallel dipoles with spacing $d$, Bound 3 (left $y$-axis) is compared with Bound 1 (right $y$-axis). Bound 3 is the same as the black curve shown in Figure 3. The dotted horizontal line indicates the limiting values for decoupled dipoles.

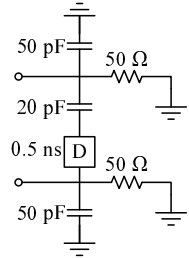

Fig. 5. Two capacitively coupled loads. The coupling includes an ideal delay component of 0.5 ns.

In order to apply the broadband bounds to Figure 5, we need to approximate the delay using a rational function. We consider two ways of doing this. The first way models the time delay term $e^{-5 \times 10^{-10}s}$ using the Padé approximation. We use the fraction

$$\frac{s^4 - 2 \times 10^{10}s^3 + 1.8 \times 10^{20}s^2 - 8.4 \times 10^{29}s + 1.68 \times 10^{39}}{s^4 + 2 \times 10^{10}s^3 + 1.8 \times 10^{20}s^2 + 8.4 \times 10^{29}s + 1.68 \times 10^{39}}.$$

This fraction is then inserted in place of the time delay, thus making the rational matrix $S_L(s)$ that satisfies $S_L(\infty) = -I$, and has poles and zeros shown in Table II(a). (We omit the expression of $S_L(s)$ because of its complexity.) Compared with the true S-matrix of the loads, the rational S-matrix has an average error of $-119$ dB in its entries between 100 MHz and 400 MHz. We apply Bound 3 and obtain $1.24 \times 10^9$. Using (9) with $\tau = 0.2$, we obtain $\delta B \approx 9.64 \times 10^7$, which is 8% of the bound. Hence (6) is

$$\int_{1 \times 10^8}^{4 \times 10^8} \log \frac{1}{r'(\omega)} d\omega \leq 1.34 \times 10^9. \quad (12)$$

This bound is larger than without the delay.

A second way to model delay is to avoid the analytical Padé step, and directly fit numerical simulations using rational

(a)

| $i$ | $p_{L,i}$ | $z_{L,i}$ |
|---|---|---|
| 1 | $-6.71 \times 10^8$ | 0 |
| 2, 3 | $(-0.05 \pm 3.37j) \times 10^9$ | $\pm 3.38j \times 10^9$ |
| 4, 5 | $(-0.00 \pm 1.33j) \times 10^{10}$ | $\pm 1.33j \times 10^{10}$ |
| 6 | $-4.35 \times 10^8$ | 0 |
| 7, 8 | $(-0.16 \pm 1.25j) \times 10^9$ | $\pm 1.32j \times 10^9$ |
| 9, 10 | $(-0.02 \pm 6.81j) \times 10^9$ | $\pm 6.81j \times 10^9$ |

(b)

| $i$ | $p_{L,i}$ | $z_{L,i}$ |
|---|---|---|
| 1 | $-6.71 \times 10^8$ | 0 |
| 2, 3 | $(-0.05 \pm 3.37j) \times 10^9$ | $\pm 3.38j \times 10^9$ |
| 4, 5 | $(-0.00 \pm 1.01j) \times 10^{10}$ | $\pm 1.01j \times 10^{10}$ |
| 6 | $-4.35 \times 10^8$ | 0 |
| 7, 8 | $(-0.16 \pm 1.25j) \times 10^9$ | $\pm 1.32j \times 10^9$ |
| 9, 10 | $(-0.02 \pm 6.76j) \times 10^9$ | $\pm 6.76j \times 10^9$ |

matrices. To illustrate this, we treat the loads in Figure 5 as a black box, obtain a numerical frequency response, and use the Matrix Fitting Toolbox to fit this response. We use ten poles and ten zeros to fit the loads in 100–400 MHz, and enforce $S_L(\infty) = -I$. The resulting $S_L(s)$ has poles and zeros shown in Table II(b); the average error between the entries true S-matrix of the loads and entries of $S_L(s)$ is $-150$ dB. Bound 3 is applied to the fitted rational model, which yields $B = 1.24 \times 10^9$, and $\delta B \approx 1.75 \times 10^7$, which is 1% of the bound. The total is $B + \delta B = 1.26 \times 10^9$. This numerically fitted model result is similar to the Padé approximation result (12), illustrating the robustness of this approach.

In the next section, we further illustrate the numerical method by fitting loads that include sections of realistic transmission lines. No analytical modeling is needed.

### E. Four commercial 2.5 GHz antennas

We consider a system where $M = N = 4$, and the loads consist of two pairs of Skycross iMAT-1115 commercial antennas designed for 2.5 GHz. The antennas are shown in Figure 6(a), numbered 1 through 4 from left to right. These antennas are simulated using Ansys HFSS in 2–4 GHz, and their de-embedded S-matrix $S'_L(j\omega)$ entries are plotted in Figure 6(b). The de-embedding removes the effect of the grounded coplanar feed lines that drive the antennas in the simulation. Each antenna in Figure 6(a) is one-tenth of a wavelength away from its neighbor at 2.5 GHz and the antennas therefore couple. From Figure 6(b), we see that antennas 1 and 2 are decoupled at 2.5 GHz, but not at other frequencies; antennas 2 and 3 have significant coupling near 2.5 GHz. We wish to examine the achievable bandwidth performance of these four antennas.

We apply the modeling method in Section II to the simulated $S'_L(j\omega)$. Figure 6(b) compares the simulated $S'_L(j\omega)$ with the $S_L(s)$; the expressions for $S_L(s)$ are not shown. The maximum of any element of $S_L(j\omega) - S'_L(j\omega)$ in 2–4 GHz is -38 dB, and the average over the frequency band is -53

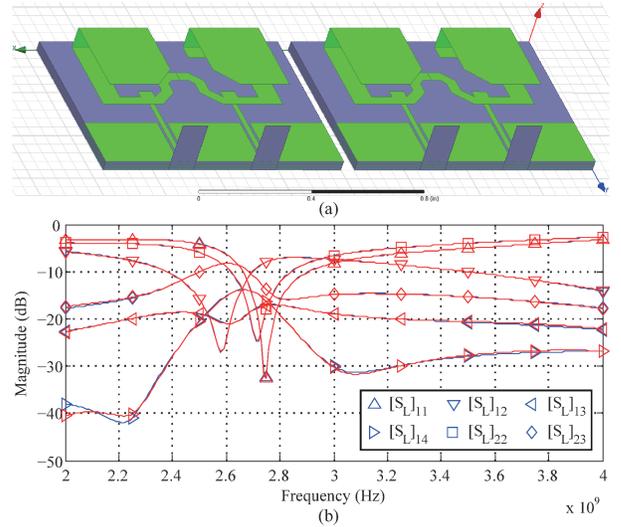

| $i$ | $p_{L,i}$ | $z_{L,i}$ |
|---|---|---|
| 1 | $-1.54 \times 10^8$ | $1.64 \times 10^8$ |
| 2 | $-7.74 \times 10^9$ | $3.67 \times 10^{10}$ |
| 3, 4 | $(-0.43 \pm 1.37j) \times 10^{10}$ | $(-0.23 \pm 2.23j) \times 10^{10}$ |
| 5, 6 | $(-0.15 \pm 1.60j) \times 10^{10}$ | $(-0.22 \pm 1.69j) \times 10^{10}$ |
| 7, 8 | $(-0.08 \pm 1.69j) \times 10^{10}$ | $(0.04 \pm 1.70j) \times 10^{10}$ |
| 9, 10 | $(-0.75 \pm 2.40j) \times 10^{10}$ | $(2.65 \pm 3.46j) \times 10^{10}$ |
| 11, 12 | $(-0.40 \pm 3.14j) \times 10^{10}$ | $(0.03 \pm 1.63j) \times 10^{10}$ |

dB. The $p_{L,i}$, $z_{L,i}$ for the antennas are listed in Table III. It is readily checked that $\det S_L(0) = 1$, and (1) is satisfied at $s_0 = 0$. Bound 1 yields

$$\int_{\omega_1}^{\omega_2} \omega^{-2} \log \frac{1}{r(\omega)} d\omega \leq 2.31 \times 10^{-10}, \quad (13)$$

where $\omega_1 = 4\pi \times 10^9$ and $\omega_2 = 8\pi \times 10^9$.

We apply the steps in Section II-B to estimate $\delta B$ for $\tau = 0.2$. The integral in (9) with $f(\omega) = 1/\omega^2$ yields $\delta B \approx 7.88 \times 10^{-11}$, which is equal to 34% of the bound in (13), and which we accept. Therefore,

$$\int_{\omega_1}^{\omega_2} \omega^{-2} \log \frac{1}{r'(\omega)} d\omega \leq B + \delta B = 3.10 \times 10^{-10} \quad (14)$$

is our bound for the four antennas.

The right-hand side of (14) is compared with the achieved bandwidth of two different matching networks. The first network is shown in Figure 7(a) for a single antenna; this network is duplicated four times, one for each antenna, and does not account for any antenna coupling. The second network, shown in Figure 7(b), is a decoupling network at 2.5 GHz designed using Method 4 in [21]. The achieved integral for the two networks are $6.98 \times 10^{-12}$ and $2.49 \times 10^{-11}$, respectively.

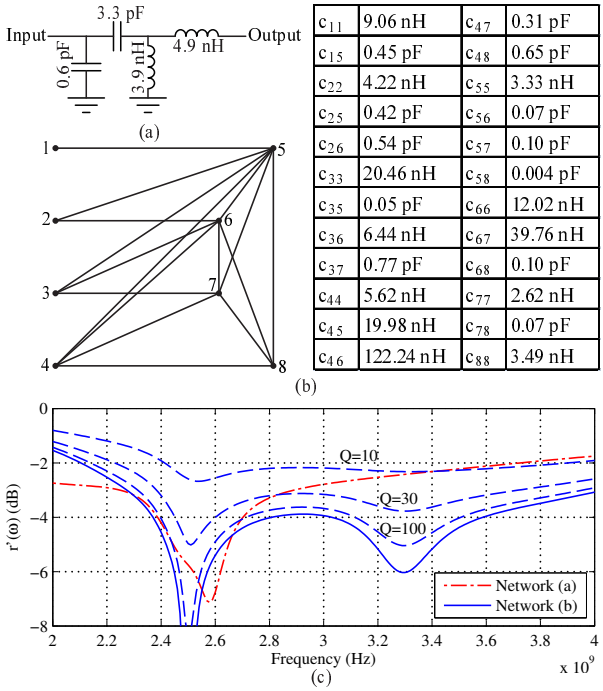

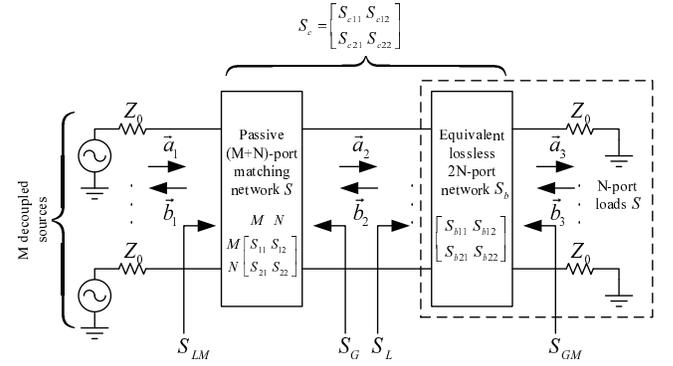

Fig. 8. An RF system where $M$ uncorrelated sources drive $N$ loads having S-matrix $S_L$ through a passive $(M+N)$-port matching network $S$ (the complex-frequency argument $s$ is omitted). The $N$ loads are shown in the dashed box using their equivalent Darlington representation, which consists of a lossless $2N$-port network $S_b$ connected with $N$ characteristic impedances $Z_0$. $S_c$ is the $(M+N) \times (M+N)$ S-matrix of the concatenated network of $S$ and $S_b$. $S_{LM}$, $S_G$ and $S_{GM}$ are the S-matrices as seen at different ports of the system.

Fig. 7. (a) A two-port matching network for a single Skycross iMAT-1115 antenna (duplicated four times) that does not account for antenna coupling. (b) A decoupling network at 2.5 GHz designed using Method 4 in [21], where each line represents a capacitor or inductor, and each port is grounded through a component not drawn in figure. The capacitance and inductance values are listed in the table to the right; $c_{ii}$ $i = 1, \ldots, 8$ are the components connecting port $i$ to ground, and $c_{ij}$ $i \neq j = 1, \ldots, 8$ are the components connecting port $i$ and $j$. (c) The $r(\omega)$ for networks (a) and (b) are shown in 2–4 GHz. Also plotted is $r(\omega)$ for (b) when the reactive components have Q factors 10, 30 and 100.

Although the decoupling network has a better bandwidth performance than the network in Figure 7(a), there still exists a significant gap between the achieved integrals and (14), indicating that much better bandwidth performance with these antennas is still possible. The $r'(\omega)$ for both networks is plotted in Figure 7(c), where we see narrow bandwidth for both networks. Neither network comes close to achieving the desired threshold of $\tau = 0.2$ (power reflection ratio $-14$ dB) over 2–4 GHz, which should be possible according to (14). The design of a network to achieve (14) remains an open problem.

The reactive components in Figure 7(b) are ideal in that they have no resistive properties. To illustrate the bandwidth lost when the components have finite Q factors, we let the reactive components behave as ideal reactive components in parallel with resistances. All of the components in Figure 7(b) are assumed to have the same Q factors for all frequency, where Q is defined as the ratio between the susceptance and the conductance of the components. Figure 7(c) shows the $r'(\omega)$ for different Q factors.

Finally, we show that the grounded coplanar waveguide feed lines used for these antennas have minimal effect on the bound. We modify the S-matrix generated by Ansys HFSS by not de-embedding it from the transmission lines, thus including them in the frequency analysis. This S-matrix is then subjected to standard rational fitting: We employ the Matrix Fitting Toolbox to fit the antennas using a rational matrix with 15 poles and zeros, and enforce (1) at $s_0 = 0$. We omit the details, but the result of Bound 1 is $B = 2.74 \times 10^{-10}$. From (9), $\delta B \approx 3.61 \times 10^{-11}$, and the result is $B + \delta B = 3.10 \times 10^{-10}$, which coincides with (14). Hence, the feed lines in this example do not affect the bandwidth, likely because they are equal in length and have minimal coupling between them.

We now provide derivations of the bounds used in both Parts of this paper.

## IV. Derivations of Broadband Matching Bounds

The bounds are corollaries of Theorems 2 and 3, which are presented in Section IV-B. The presentation of the theorems requires the Darlington representation for multiport loads, which is itself presented in Section IV-A. The bounds are derived in Section IV-C.

### A. Darlington equivalent network

We consider the RF system shown in Figure 8, where $M$ uncorrelated sources drive $N$ loads through a passive $(M + N)$-port matching network. The network has an $(M + N) \times (M + N)$ S-matrix $S(s)$ partitioned as

$$S(s) = \begin{matrix} & M & N \\ \begin{matrix} M \\ N \end{matrix} & \begin{pmatrix} S_{11}(s) & S_{12}(s) \\ S_{21}(s) & S_{22}(s) \end{pmatrix} \end{matrix}.$$

In Figure 8, we transform the $N$ dissipative real loads $S_L(s)$ into an equivalent lossless real $2N$-port network $S_b(s)$ terminated by $N$ isolated characteristic impedances $Z_0$. In [20], Darlington first verified that such an equivalent transformation is possible for $N = 1$. The extension to $N > 1$ is shown in [23]–[26]. The fact that $I - S_L^T(-s)S_L(s)$ has full normal rank ensures that there are $N$ resistors in the Darlington network [25, III.3.1].

We partition the $2N \times 2N$ S-matrix $S_b(s)$ as

$$S_b(s) = \begin{bmatrix} S_{b11}(s) & S_{b12}(s) \\ S_{b21}(s) & S_{b22}(s) \end{bmatrix},$$

where $S_{bij}(s)$ are $N \times N$ submatrices. Then the necessary and sufficient condition for $S_b(s)$ being a Darlington equivalent network for a real-rational $S_L(s)$ is that $S_b(s)$ is real-rational, Hurwitzian, bounded and para-unitary, and $S_{b11}(s) = S_L(s)$. We do not need to know the exact form of the rest of $S_b(s)$, only its existence is needed.

In Figure 8, let $S_c(s)$ be the $(M+N) \times (M+N)$ S-matrix of the concatenated network of $S(s)$ and $S_b(s)$, partitioned as

$$S_c(s) = \begin{array}{c} M \\ N \end{array} \begin{pmatrix} \overset{M}{S_{c11}(s)} & \overset{N}{S_{c12}(s)} \\ S_{c21}(s) & S_{c22}(s) \end{pmatrix}.$$

Let $S_{GM}(s)$ be the $N \times N$ S-matrix seen from the output ports of $S_b(s)$, and $S_G(s)$ be the $N \times N$ S-matrix seen from the output ports of $S(s)$. It follows that $S_G(s) = S_{22}(s)$, $S_{GM}(s) = S_{c22}(s)$, and

$$S_{GM}(s) = S_{b22}(s) + S_{b21}(s)S_G(s)(I - S_L(s)S_G(s))^{-1}S_{b12}(s). \quad (15)$$

We use the notation $p_{\times,i}$ and $z_{\times,i}$, $i = 1, 2, \ldots$ to represent the poles and zeros over the WCP of $S_\times$, where $S_\times$ is any of the S-matrices or submatrices shown in Figure 8. In addition, we use the subscript "+" to denote the zeros or poles that are in the RHP, and "−" to denote those in the LHP.

In Figure 8, let $\vec{a}_3(s)$ and $\vec{b}_3(s)$ be the $N \times 1$ incident and reflected signal to and from the isolated impedances $Z_0$. Because $S_b(s)$ is lossless, the amount of power delivered to the loads equals the power delivered to the resistive part of the Darlington network. Since $\vec{b}_3(j\omega) = 0$, the power delivered at $j\omega$ is $\|\vec{a}_3(j\omega)\|^2$, where $\vec{a}_3(j\omega) = S_{c21}(j\omega)\vec{a}_1(j\omega)$. So the total power lost to reflection and dissipation can be written as $\|\vec{a}_1(j\omega)\|^2 - \|\vec{a}_3(j\omega)\|^2$, and the power loss ratio of the matching network and the loads becomes

$$r^2(\omega) = 1 - \frac{1}{M} \mathrm{tr}\{S_{c21}^H(j\omega)S_{c21}(j\omega)\}. \quad (16)$$

In the rest of the presentation, we use a particular Darlington network, which is given in the following lemma in [9].

*Lemma 1:* There exists a Darlington network $S_b(s)$ such that

$$z_{b22+,i} = -z_{L-,i}. \quad (17)$$

For such $S_b(s)$, the RHP zeros of $S_{GM}(s)$ are identical to the RHP zeros of $S_L^T(-s) - S_G(s)$.

### B. Integral log-determinant of $S_{GM}(j\omega)$

We now present the theorems on the integral of logarithm of $\det S_{GM}(j\omega)$. We assume that $I - S_L^T(-s)S_L(s)$ is full normal rank. We also assume (1) is satisfied for some $\mathrm{Re}\{s_0\} \geq 0$, and choose positive integer $m$ such that

$$I - S_L^T(-s)S_L(s) = A_m(s - s_0)^m + A_{m+1}(s - s_0)^{m+1} + \ldots \quad (18)$$

if $s_0$ is finite, or

$$I - S_L^T(-s)S_L(s) = A_m s^{-m} + A_{m+1}s^{-(m+1)} + \ldots \quad (19)$$

if $s_0 = \infty$. In (18) and (19), $A_m \neq 0$ is defined such that the entries of $I - S_L^T(-s)S_L(s)$ have a zero at $s = s_0$ with multiplicity at least $m$. When $s_0 = j\omega_0$, it is shown in Lemma 4 in Appendix C that $m$ is even; when $s_0 = \infty$, $m$ is also even since the left-hand side of (19) is an even function. The general broadband matching theorems are as follows:

*Theorem 2:* Let $S_L(s)$ satisfy (18) for some $\mathrm{Re}\{s_0\} \geq 0$. Then for any passive network $S(s)$ such that $I - S_L(s_0)S_G(s_0)$ is non-singular, we have

$$\int_0^\infty \mathrm{Re}[(s_0 - j\omega)^{-1} + (s_0 + j\omega)^{-1}] \log|\det S_{GM}(j\omega)|d\omega$$
$$= \pi \log\left| \det S_L(s_0) \cdot \frac{\prod_i(s_0 + z_{L,i})\prod_i(s_0 + z_{GM+,i})}{\prod_i(s_0 - z_{L,i})\prod_i(s_0 - z_{GM+,i})} \right|, \quad (20)$$

and

$$\int_0^\infty [(s_0 - j\omega)^{-(k+1)} + (s_0 + j\omega)^{-(k+1)}] \times$$
$$\log|\det S_{GM}(j\omega)|d\omega = \frac{(-1)^k\pi}{k}\Big[ \sum_i(p_{L,i} - s_0)^{-k}$$
$$- \sum_i(-z_{L,i} - s_0)^{-k} + \Big(\sum_i(z_{GM+,i} - s_0)^{-k}$$
$$- \sum_i(-z_{GM+,i} - s_0)^{-k}\Big)\Big] \quad (21)$$

for $k = 1, \ldots, m-1$, where $\pm s_0$, $\pm s_0^*$ are excluded in $p_{L,i}$, $z_{L,i}$ in (21). If $I - S_L(s_0)S_G(s_0)$ is singular, then $\mathrm{Re}\{s_0\} = 0$, (20) holds, and (21) holds for $k = 1, \ldots, m-2$; for $k = m-1$, we let $s_0 = j\omega_0$, and then have

$$\int_0^\infty [(\omega_0 - \omega)^{-m} + (\omega_0 + \omega)^{-m}] \log|\det S_{GM}(j\omega)|d\omega$$
$$\geq \frac{(-1)^{\frac{m-2}{2}}\pi}{m-1}\Big[ \sum_i(p_{L,i} - j\omega_0)^{-(m-1)}$$
$$- \sum_i(-z_{L,i} - j\omega_0)^{-(m-1)} + \Big(\sum_i(z_{GM+,i} - j\omega_0)^{-(m-1)}$$
$$- \sum_i(-z_{GM+,i} - j\omega_0)^{-(m-1)}\Big)\Big]. \quad (22)$$

*Theorem 3:* Let $S_L(s)$ satisfy (19). Then

$$\int_0^\infty \omega^{k-1} \log|\det S_{GM}(j\omega)|d\omega$$
$$\geq \frac{(-1)^{\frac{k-1}{2}}\pi}{2k}\Big[ \sum_i p_{L,i}^k + \sum_i z_{L,i}^k + 2\sum_i z_{GM+,i}^k \Big] \quad (23)$$

for $k = 1, 3, \ldots, m-1$. Equality in (23) holds if $k \neq m-1$ or $I - S_L(s)S_G(s)$ is non-singular at $s = \infty$.

In Theorem 2, although (18) holds, $s_0$ may still be a zero of $S_L(s)$ when $\mathrm{Re}\{s_0\} > 0$. This happens if and only if $S_L(s)$ has a pole at $-s_0$, which cancels the zero at $s_0$ in (18). Hence, the poles and zeros of $S_L(s)$ at $\pm s_0$, $\pm s_0^*$ are excluded in the sums in (21). The proof of Theorem 2 appears in Appendix D; the proof uses several preliminary lemmas which are introduced in Appendices B and C.

Theorem 3 is adapted from Theorem 1 in [9], which is proven with $\det S_{LM}(j\omega)$ in place of $\det S_{GM}(j\omega)$ in (23). Note $S_{LM}(s)$ is the $M \times M$ S-matrix seen from the input ports of $S(s)$ (see Figure 8). Although it is assumed in [9] that $M = N$, and that the matching network is lossless and reciprocal, and the loads are also reciprocal, these assumptions can be relaxed. In fact, Theorem 1 in [9] applies without change to non-reciprocal networks and loads; the reciprocity of the networks and loads is an unnecessary restriction in the model. Our version is needed to handle lossy networks, which are not handled in [9]. The proof of (23) follows the same arguments used in the proof of Theorem 2 and is omitted.

### C. Proof of bounds

We relate $\det S_{GM}(j\omega)$ to $r(\omega)$ in (16) using the arithmetic-geometric mean inequality:

$$
\begin{aligned}
r^2(\omega) &= \frac{1}{M}\mathrm{tr}\{I - S_{c21}^H(j\omega)S_{c21}(j\omega)\} \\
&\geq \det(I - S_{c21}(j\omega)S_{c21}^H(j\omega))^{1/M} \\
&\geq \det(S_{GM}(j\omega)S_{GM}^H(j\omega))^{1/M} = |\det S_{GM}(j\omega)|^{2/M}.
\end{aligned}
$$

The first equality holds if and only if the eigenvalues of $I - S_{c21}^H(j\omega)S_{c21}(j\omega)$ are all equal; this is equivalent to $S_{21}^H(I - S_G S_L)^{-H}(I - S_H^H S_L)(I - S_G S_L)^{-1}S_{21}$ having equal singular values for all $j\omega$, which is Condition 2 for equality in [1]. The second equality holds if and only if $S(s)$ satisfies $S_{c21}(s)S_{c21}^T(-s) + S_{GM}(s)S_{GM}^T(-s) = I$. Since $S_b(s)$ is para-unitary, this is equivalent to $S_{21}(s)S_{21}^T(-s) + S_G(s)S_G^T(-s) = I$, which is Condition 1.

Taking the logarithm on both sides of the inequality yields

$$\log r(\omega) \geq (1/M)\log|\det S_{GM}(j\omega)|. \tag{24}$$

When $s_0 = j\omega_0$, we apply (24) to (21) and (22) for $k = 1$, and omit $\sum_i (z_{GM+,i} - j\omega_0)^{-1} - \sum_i (-z_{GM+,i} - j\omega_0)^{-1}$ since it is non-negative. From Lemma 1, the RHP zeros of $S_{GM}(s)$ are identical to the RHP zeros of $S_L^T(-s) - S_G(s)$; thus Condition 4 is necessary and one of the sufficient conditions. This finishes the proof of Bound 1.

When $\mathrm{Re}\{s_0\} > 0$, we apply (24) to (20), and then omit $\frac{\prod_i(s_0 + z_{GM+,i})}{\prod_i(s_0 - z_{GM+,i})}$ since it has modulus no smaller than one. Note $\mathrm{Re}[(s_0 - j\omega)^{-1} + (s_0 + j\omega)^{-1}]$ in (20) is positive for any $\mathrm{Re}\{s_0\} > 0$ and $\omega$. The result is Bound 2.

We combine (24) and (23) for $k = 1$, and then omit $\sum_i z_{GM+,i}$ since it is non-negative. The result is Bound 3. ∎

## V. Conclusions and Future Work

We have presented a bandwidth analysis for multiport matching that applies to an arbitrary number of sources and coupled loads, using broadband matching bounds on the integral of a power loss ratio. Part I presented the definitions and bounds and applied them to settings where the loads could be described analytically using rational functions. Conditions were given for when the bounds could be met with equality. Part II focused on realistic loads, including antennas, and the rational fitting needed for accurate bandwidth calculations. Proofs of all results were also presented.

Part I demonstrated that the bound scales generally as $N/M$ for $M$ sources and $N$ loads. This scaling is not affected by coupling as long as it is not "too strong". Hence, large bandwidth is possible even if the loads are coupled or closely-spaced. Some realistic examples in Part II showed how large bandwidths can be attained in practice. Although we touched upon techniques to attain the bounds, the general practical design problem remains open.

For loads, such as antennas, whose frequency responses are available numerically through simulation, rational fitting of the S-matrix is required to apply the bounds. We showed how the rational matrix is then used to find a bandwidth bound for the system being fitted. Increasing the polynomial orders of the numerators and denominators of the rational functions in the S-matrix generally improves the accuracy of the system approximation, but can cause the problem of over-fitting, which results in a loose bound. Over-fitting is an important issue that deserves further attention.

Our examples included antennas that are matched near their resonant frequency. It would be interesting to see how the bounds characterize the bandwidth for electrically small antennas, which are being used below their resonant frequencies and are generally considered narrowband. The classical Chu bound [28] and some recent advances [29]–[31] provide other methods to describe the bandwidth for small antennas. One future area of study could be to reconcile these methods with bandwidth results calculated using the modeling techniques and bounds contained herein.

The communication-theoretic implications of using broadband multiport networks in wireless systems need study. We have been concerned primarily with the aspects of network design that ensure efficient power delivery to loads such as an array of multiple coupled antennas. However, the choice of network can also affect the radiation efficiency and far-field pattern of an antenna array. Hence, a complete system design should consider total power transfer from the transmitter amplifiers to a far-field receiver. We also did not examine the implications of using the wideband matching networks for antennas that are intended for both transmission and reception.

Although we have treated the case where uncorrelated sources are driving coupled loads, the reverse situation where coupled sources are driving decoupled loads needs separate analysis. Such a system could arise when the sources are closely-spaced receiver antennas connected to isolated low-noise amplifiers. The noise figures of the amplifiers would play a role in the criterion for determining bandwidth.

Maximizing the data rate attainable with a prescribed set of $N$ antennas is a potentially interesting problem. Conventional narrowband Shannon theory says that MIMO, with $M$ independent data streams (sources) in a rich scattering environment, achieves rates linear in $N$ when $M = N$ [32]. But, as we have seen, the bandwidth attainable for $M = N$ is only $1/N$ of that attainable for $M = 1$. Thus, MIMO with $N$ streams achieves $1/N$ the bandwidth of MIMO with a single stream. Since the transmission data rate is directly proportional to bandwidth, one could therefore conjecture that, with $N$ antennas, one stream is as good as $N$. This bandwidth/multiplexing trade-off needs further study.

## VI. Acknowledgements

We thank the reviewers and editor for the detailed and insightful comments on both parts of our manuscript. In particular, your technical suggestions on how to improve the readability of the paper helped us to treat the issues of the experimental determination of $r(\omega)$, and the rational fitting of realistic loads.

## Appendix

Theorem 1 is proven in Appendix A. We prove Theorem 2 in Appendix D using several preliminary lemmas that are introduced in Appendices B and C.

### A. Proof of Theorem 1

Let $\varepsilon(\omega) = r^2(\omega) - r'^2(\omega)$. Then

$$\int_{\omega_1}^{\omega_2} f(\omega) \log \frac{r(\omega)}{r'(\omega)} d\omega = \int_{\omega_1}^{\omega_2} \frac{f(\omega)}{2} \log\left(1 + \frac{\varepsilon(\omega)}{r'^2(\omega)}\right) d\omega$$
$$= \int_{\omega_1}^{\omega_2} \frac{f(\omega)}{2} \log\left(1 + \frac{1 - r^2(\omega)}{r'^2(\omega)} \cdot \frac{\varepsilon(\omega)}{1 - r^2(\omega)}\right) d\omega.$$

We show that $\varepsilon(\omega)/(1 - r'^2(\omega)) \leq \rho(\omega)$.

From (5) in Part I,

$$r^2(\omega) = 1 - \frac{\operatorname{tr}\{S_{21}^H(I - S_G S_L)^{-H}(I - S_L^H S_L)(I - S_G S_L)^{-1} S_{21}\}}{M},$$

and a similar relation between $r'^2(\omega)$ and $S_L'(j\omega)$ holds. For simplicity of the presentation, we omit the argument $j\omega$ for the S-matrices in the remainder of the proof. We perturb $S_L(j\omega) = S_L'(j\omega) + \delta S_L(j\omega)$; a first-order expansion yields

$$\varepsilon(\omega) = r^2(\omega) - r'^2(\omega) \approx$$
$$-\frac{1}{M} \operatorname{tr}\{S_{21}^H(I - S_G S_L')^{-H} A_1 (I - S_G S_L')^{-1} S_{21}\},$$

where $A_1$ is the Hermitian matrix

$$A_1 = \delta S_L^H (I - S_L' S_G)^{-H}(S_G^H - S_L')$$
$$+ (S_G - S_L'^H)(I - S_L' S_G)^{-1} \delta S_L.$$

Because $A_1$ is Hermitian, the matrix $S_{21}^H(I - S_G S_L')^{-H} A_1 (I - S_G S_L')^{-1} S_{21}$ is also Hermitian. Some manipulations of the trace give us

$$\varepsilon(\omega) \leq \frac{\sigma_{1,\max}}{M} \operatorname{tr}\{S_{21}^H(I - S_G S_L')^{-H}(I - S_G S_L')^{-1} S_{21}\},$$

where $\sigma_{1,\max}$ is the maximum singular value of $A_1$. Similarly,

$$1 - r'^2(\omega)$$
$$= \frac{1}{M} \operatorname{tr}\{S_{21}^H(I - S_G S_L')^{-H}(I - S_L'^H S_L')(I - S_G S_L')^{-1} S_{21}\}$$
$$\geq \frac{1 - \sigma_{L,\max}'^2(\omega)}{M} \operatorname{tr}\{S_{21}^H(I - S_G S_L')^{-H}(I - S_G S_L')^{-1} S_{21}\}.$$

Hence,

$$\frac{\varepsilon(\omega)}{1 - r'^2(\omega)} \leq \frac{\sigma_{1,\max}}{1 - \sigma_{L,\max}'^2(\omega)}.$$

We now seek an upper bound on $\sigma_{1,\max}$, which depends on the maximum singular values of both $\delta S_L$ and $(S_G - S_L'^H)(I - S_L' S_G)^{-1}$. The maximum singular value of $\delta S_L$ is $\sigma_{\delta,\max}(\omega)$. To obtain the maximum singular value of $(S_G - S_L'^H)(I - S_L' S_G)^{-1}$, we try to prove that

$$I + (\sigma_{L,\max}'^2(\omega) - \sigma_{L,\min}'^2(\omega))(I - S_L' S_G)^{-H}(I - S_L' S_G)^{-1}$$
$$- (I - S_L' S_G)^{-H}(S_G^H - S_L')(S_G - S_L'^H)(I - S_L' S_G)^{-1} \quad (25)$$

is a positive semidefinite matrix, which means the maximum eigenvalue of the positive definite matrix $I + (\sigma_{L,\max}'^2(\omega) - \sigma_{L,\min}'^2(\omega))(I - S_L' S_G)^{-H}(I - S_L' S_G)^{-1}$ is larger than or equal to the maximum eigenvalue of the positive semidefinite matrix $(I - S_L' S_G)^{-H}(S_G^H - S_L')(S_G - S_L'^H)(I - S_L' S_G)^{-1}$. Then the square root of the maximum eigenvalue of $I + (\sigma_{L,\max}'^2(\omega) - \sigma_{L,\min}'^2(\omega))(I - S_L' S_G)^{-H}(I - S_L' S_G)^{-1}$ is larger than or equal to the maximum singular value of $(S_G - S_L'^H)(I - S_L' S_G)^{-1}$.

To prove that (25) is positive semidefinite, we simplify it

$$I + (\sigma_{L,\max}'^2(\omega) - \sigma_{L,\min}'^2(\omega))(I - S_L' S_G)^{-H}(I - S_L' S_G)^{-1}$$
$$- (I - S_L' S_G)^{-H}(S_G^H - S_L')(S_G - S_L'^H)(I - S_L' S_G)^{-1}$$
$$= (I - S_L' S_G)^{-H} A_2 (I - S_L' S_G)^{-1},$$

where

$$A_2 = (\sigma_{L,\max}'^2(\omega) - \sigma_{L,\min}'^2(\omega))I + (I - S_L' S_L'^H)$$
$$- S_G^H(I - S_L'^H S_L')S_G.$$

$A_2$ is a positive semidefinite matrix because the singular values of $S_L'$ and $S_G$ are no larger than one, and therefore the minimum eigenvalue of the positive definite matrix $(\sigma_{L,\max}'^2(\omega) - \sigma_{L,\min}'^2(\omega))I + (I - S_L' S_L'^H)$ is larger than or equal to the maximum eigenvalue of the positive semidefinite matrix $S_G^H(I - S_L'^H S_L')S_G$. Therefore, we have proven that the square root of the maximum eigenvalue of $I + (\sigma_{L,\max}'^2(\omega) - \sigma_{L,\min}'^2(\omega))(I - S_L' S_G)^{-H}(I - S_L' S_G)^{-1}$ is larger than or equal to the maximum singular value of $(S_G - S_L'^H)(I - S_L' S_G)^{-1}$. The maximum eigenvalue of $I + (\sigma_{L,\max}'^2(\omega) - \sigma_{L,\min}'^2(\omega))(I - S_L' S_G)^{-H}(I - S_L' S_G)^{-1}$ is smaller than or equal to $1 + \frac{\sigma_{L,\max}'^2(\omega) - \sigma_{L,\min}'^2(\omega)}{(1 - \sigma_{L,\max}'(\omega))^2}$. Therefore, the maximum singular value of $(S_G - S_L'^H)(I - S_L' S_G)^{-1}$ is smaller than or equal to $\sqrt{1 + \frac{\sigma_{L,\max}'^2(\omega) - \sigma_{L,\min}'^2(\omega)}{(1 - \sigma_{L,\max}'(\omega))^2}}$. We conclude

$$\sigma_{1,\max} \leq 2\sigma_{\delta,\max}(\omega) \sqrt{1 + \frac{\sigma_{L,\max}'^2(\omega) - \sigma_{L,\min}'^2(\omega)}{(1 - \sigma_{L,\max}'(\omega))^2}},$$

which yields $\varepsilon(\omega)/(1 - r'^2(\omega)) \leq \rho(\omega)$. This finishes the proof. ∎

## B. Preliminary lemmas on real-rational functions

*Lemma 2:* Let $f(s)$ be a real-rational function with $f(s_0) = c \neq 0$. Then the series expansion of $\log f(s)$ around $s = s_0$ can be written as

$$
\log f(s) = \log c + a_1(s - s_0) + a_2(s - s_0)^2 + \ldots \\
+ a_\ell (s - s_0)^\ell + \ldots, \tag{26}
$$

where

$$
a_\ell = \frac{1}{\ell} \left( \sum_i (p_i - s_0)^{-\ell} - \sum_i (z_i - s_0)^{-\ell} \right) \tag{27}
$$

for $\ell = 1, 2, \ldots$, where $p_i, z_i$ are the poles and zeros of $f(s)$.

*Proof:* Since $f(s_0) \neq 0$ is finite, $\log f(s_0) = \log c$ is finite. So the expansion of $\log f(s)$ around $s = s_0$ can be obtained using Taylor series expansion. The result is (27). ■

*Lemma 3:* Let $f(s)$ be a real-rational function. For any $\mathrm{Re}\{s_0\} \geq 0$, if $f(s_0)$ is finite, non-zero, and $1 - f(-s)f(s)$ has a zero at $s = s_0$ with multiplicity $m$, then

$$
\int_0^\infty \mathrm{Re}[(s_0 - j\omega)^{-1} + (s_0 + j\omega)^{-1}] \log |f(j\omega)| d\omega \\
= \pi \log \left| f(s_0) \frac{\prod_i (s_0 - p_{+,i}) \prod_i (s_0 + z_{+,i})}{\prod_i (s_0 + p_{+,i}) \prod_i (s_0 - z_{+,i})} \right|, \tag{28}
$$

and

$$
\int_0^\infty [(s_0 - j\omega)^{-(k+1)} + (s_0 + j\omega)^{-(k+1)}] \log |f(j\omega)| d\omega \\
= \frac{(-1)^k \pi}{k} \Big[ \sum_i (p_i - s_0)^{-k} - \sum_i (z_i - s_0)^{-k} \\
- \Big( \sum_i (p_{+,i} - s_0)^{-k} - \sum_i (-p_{+,i} - s_0)^{-k} \Big) \\
+ \Big( \sum_i (z_{+,i} - s_0)^{-k} - \sum_i (-z_{+,i} - s_0)^{-k} \Big) \Big] \tag{29}
$$

for $k = 1, \ldots, m-1$, where $p_i, z_i$ are the poles and zeros of $f(s)$ in the WCP, and $p_{+,i}, z_{+,i}$ are the poles and zeros of $f(s)$ in the RHP.

*Proof:* Since $f(s_0) \neq 0$, we begin by applying Lemma 2 to write the expansion of $\log f(s)$ as (26) and (27). For convenience, we define another real-rational function $\hat{f}(s)$ as

$$
\hat{f}(s) = f(s) \frac{\prod_i (s - p_{+,i}) \prod_i (s + z_{+,i})}{\prod_i (s + p_{+,i}) \prod_i (s - z_{+,i})}, \tag{30}
$$

where $p_{+,i}$ and $z_{+,i}$ are the poles and zeros of $f(s)$ in the RHP. Then $\hat{f}(s)$ has no poles or zeros in the RHP, and satisfies $\hat{f}(-s)\hat{f}(s) = f(-s)f(s)$ and $|\hat{f}(j\omega)| = |f(j\omega)|$. We apply Lemma 2 to $\hat{f}(s)$ and get

$$
\log \hat{f}(s) = \log \hat{c} + \hat{a}_1(s - s_0) + \hat{a}_2(s - s_0)^2 + \ldots \\
+ \hat{a}_\ell (s - s_0)^\ell + \ldots, \tag{31}
$$

where

$$
\hat{c} = \hat{f}(s_0) = f(s_0) \frac{\prod_i (s_0 - p_{+,i}) \prod_i (s_0 + z_{+,i})}{\prod_i (s_0 + p_{+,i}) \prod_i (s_0 - z_{+,i})}, \tag{32}
$$

and

$$
\hat{a}_\ell = a_\ell - \frac{1}{\ell} \left( \sum_i (p_{+,i} - s_0)^{-\ell} - \sum_i (-p_{+,i} - s_0)^{-\ell} \right)
$$

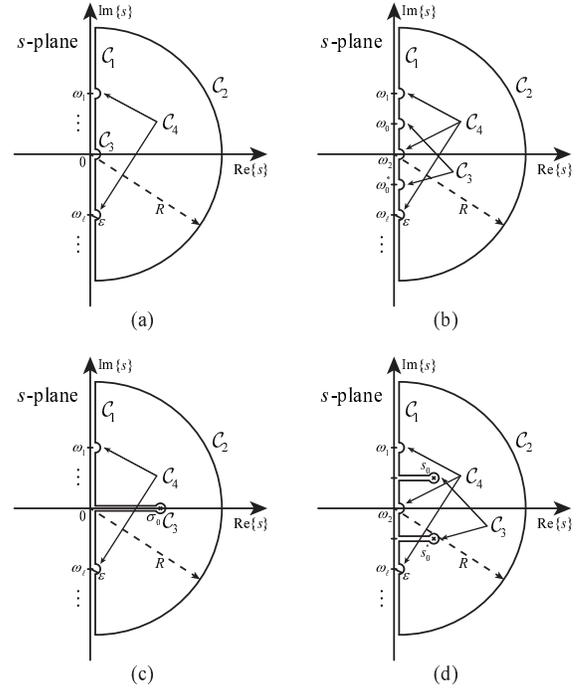

Fig. 9. The contours for the integrals (35), (37), (39) and (41) are shown in (a)–(d), respectively. The contours are in the clockwise direction. The sections of the contours are labeled $\mathcal{C}_1, \mathcal{C}_2, \ldots$ and their detailed descriptions are given in the proof of Lemma 3.

$$
+ \frac{1}{\ell} \left( \sum_i (z_{+,i} - s_0)^{-\ell} - \sum_i (-z_{+,i} - s_0)^{-\ell} \right). \tag{33}
$$

We can expand $\log f(s)$ and $\log \hat{f}(s)$ at $s = s_0^*$ in similar forms as (26) and (31). Since $f(s)$ and $\hat{f}(s)$ are real-rational, the coefficients for the expansion of $\log f(s)$ and $\log \hat{f}(s)$ at $s = s_0^*$ are $a_\ell^*$ and $\hat{a}_\ell^*$, respectively.

The next step is to separate our discussions into four cases: $s_0 = 0$, $j\omega_0$, $\sigma_0$, $\sigma_0 + j\omega_0$. Since the cases are similar to one another, we elaborate more on the $s_0 = 0$ case than the others.

*1) $s_0 = 0$:* Since $1 - f(-s)f(s)$ is an even real-rational function and has a zero of multiplicity $m$ at $s_0 = 0$, $m$ must be an even integer. We take the logarithm of $\hat{f}(-s)\hat{f}(s) = f(-s)f(s) = 1 + O(s^m)$ and use the expansion (31) at $s_0 = 0$. Because $\hat{f}(s)$ is real-rational, the coefficients in (33) for $s_0 = 0$ are real. We therefore obtain $|\hat{c}| = 1$, $\hat{a}_2 = \hat{a}_4 = \ldots = \hat{a}_{m-2} = 0$ and $\mathrm{Im}\{\hat{a}_1\} = \mathrm{Im}\{\hat{a}_3\} = \ldots = \mathrm{Im}\{\hat{a}_{m-1}\} = 0$.

For $s_0 = 0$, (28) is trivial since both sides are zero. Because $\hat{a}_2 = \hat{a}_4 = \ldots = \hat{a}_{m-2} = 0$, (29) is also trivial for $k = 2, 4, \ldots, m-2$. We show (29) for $k = 1, 3, \ldots, m-1$ by taking the contour integral of the function

$$
s^{-(k+1)} \log \hat{f}(s) \tag{34}
$$

along the closed curve shown in Figure 9(a). This function is analytic in the RHP; it is also analytic on the imaginary axis except the origin and possible zeros and poles of $\hat{f}(s)$ on the imaginary axis, which we denote as $j\omega_\ell$. Therefore, the following contour integral is zero:

$$
\int_{\mathcal{C}_1 + \mathcal{C}_2 + \mathcal{C}_3 + \mathcal{C}_4} s^{-(k+1)} \log \hat{f}(s) ds = 0, \tag{35}
$$

where $\mathcal{C}_1$ is the line segment between $-jR$ and $jR$ excluding $[-\varepsilon, \varepsilon]$ and $[j(\omega_\ell - \varepsilon), j(\omega_\ell + \varepsilon)]$; $\mathcal{C}_2$ is the right semicircle with radius $R$ centered at the origin; $\mathcal{C}_3$ is the right semicircle with radius $\varepsilon$ centered at the origin; and $\mathcal{C}_4$ includes the right semicircles with radius $\varepsilon$ centered at $j\omega_\ell$. We evaluate the integral of (34) as follows:

$$\int_{\mathcal{C}_1} s^{-(k+1)} \log \hat{f}(s) ds$$
$$= (-1)^{\frac{k+1}{2}} j \int_{-R}^{R} \omega^{-(k+1)} \log |f(j\omega)| d\omega - \frac{(-1)^{\frac{k+1}{2}} 2 \arg(\hat{c})}{k \varepsilon^k}$$
$$+ O(R^{-k}),$$

where the integral from $-R$ to $R$ excludes the intervals $[-\varepsilon, \varepsilon]$ and $[\omega_\ell - \varepsilon, \omega_\ell + \varepsilon]$. Furthermore,

$$\int_{\mathcal{C}_2} s^{-(k+1)} \log \hat{f}(s) ds = \int_{\pi/2}^{-\pi/2} j(Re^{j\theta})^{-k} \log \hat{f}(Re^{j\theta}) d\theta$$
$$= O\left(\frac{\log R}{R}\right).$$

$$\int_{\mathcal{C}_3} s^{-(k+1)} \log \hat{f}(s) ds = \int_{-\pi/2}^{\pi/2} j(\varepsilon e^{j\theta})^{-k} \log \hat{f}(\varepsilon e^{j\theta}) d\theta$$
$$= j\pi \hat{a}_k + \frac{(-1)^{\frac{k+1}{2}} 2 \arg(\hat{c})}{k \varepsilon^k} + O(\varepsilon^{m-k}).$$

$$\int_{\mathcal{C}_4} s^{-(k+1)} \log \hat{f}(s) ds = \sum_{\ell} \int_{-\pi/2}^{\pi/2} (j\omega_\ell + \varepsilon e^{j\theta})^{-(k+1)}$$
$$\times \log \hat{f}(j\omega_\ell + \varepsilon e^{j\theta}) j\varepsilon e^{j\theta} d\theta = O(\varepsilon \log \varepsilon).$$

Combining these path integrals and letting $R \to \infty$, $\varepsilon \to 0$, we have

$$\int_{-\infty}^{\infty} \omega^{-(k+1)} \log |f(j\omega)| d\omega = (-1)^{\frac{k-1}{2}} \pi \hat{a}_k.$$

Since $f(s)$ is real-rational, $|f(-j\omega)| = |f^*(j\omega)| = |f(j\omega)|$, we have $\omega^{-(k+1)} \log |f(j\omega)|$ is an even function for $k = 1, 3, \ldots, m-1$. Using (33), we get (29) for $k = 1, 3, \ldots, m-1$. This finishes the proof for $s_0 = 0$.

*2) $s_0 = j\omega_0$:* Since $1 - f(-s)f(s)$ is a real-rational function and has a zero of multiplicity $m$ at $s = j\omega_0$, it also has a zero of multiplicity $m$ at $s = -j\omega_0$. We take the logarithm of $\hat{f}(-s)\hat{f}(s) = f(-s)f(s) = 1 + O((s - j\omega_0)^m)$ and use the expansion (31) at $s_0 = j\omega_0$. We therefore obtain $|\hat{c}| = 1$, $\text{Im}\{a_1\} = \text{Im}\{a_3\} = \ldots = \text{Im}\{a_k\} = 0$ for odd $k$ and $k < m$, and $\text{Re}\{a_2\} = \text{Re}\{a_4\} = \ldots = \text{Re}\{a_k\} = 0$ for even $k$ and $k < m$.

For $s_0 = j\omega_0$, (28) is trivial since both sides are zero. We show (29) by taking the contour integral of the function

$$[(j\omega_0 - s)^{-(k+1)} + (j\omega_0 + s)^{-(k+1)}] \log \hat{f}(s) \qquad (36)$$

along the closed curve shown in Figure 9(b). This function is analytic in the RHP; it is also analytic on the imaginary axis except $\pm j\omega_0$ and possible zeros and poles of $\hat{f}(s)$ on the imaginary axis, which we denote as $j\omega_\ell$. Therefore,

$$\int_{\mathcal{C}_1 + \mathcal{C}_2 + \mathcal{C}_3 + \mathcal{C}_4} [(j\omega_0 - s)^{-(k+1)} + (j\omega_0 + s)^{-(k+1)}]$$

$$\times \log \hat{f}(s) ds = 0, \qquad (37)$$

where $\mathcal{C}_1$ is the line segment between $-jR$ and $jR$ excluding $[j(\omega_0 - \varepsilon), j(\omega_0 + \varepsilon)]$, $[j(-\omega_0 - \varepsilon), j(-\omega_0 + \varepsilon)]$ and $[j(\omega_\ell - \varepsilon), j(\omega_\ell + \varepsilon)]$; $\mathcal{C}_2$ is the right semicircle with radius $R$ centered at the origin; $\mathcal{C}_3$ includes the right semicircles with radius $\varepsilon$ centered at the $j\omega_0$ and $-j\omega_0$; and $\mathcal{C}_4$ includes the right semicircles with radius $\varepsilon$ centered at $j\omega_\ell$.

By evaluating the integral paths in (37), and letting $R \to \infty$, $\varepsilon \to 0$, we obtain

$$\int_{-\infty}^{\infty} [(\omega_0 - \omega)^{-(k+1)} + (\omega_0 + \omega)^{-(k+1)}] \log |f(j\omega)| d\omega$$
$$= (-1)^k j^{k+1} 2\pi \hat{a}_k.$$

We have $[(\omega_0 - \omega)^{-(k+1)} + (\omega_0 + \omega)^{-(k+1)}] \log |f(j\omega)|$ is an even function, and using (33) we get (29) for $k = 1, 2, \ldots, m-1$. This finishes the proof for $s_0 = j\omega_0$.

*3) $s_0 = \sigma_0$:* Since $f(s)$ is a real-rational function, $\text{Im}\{\hat{a}_1\} = \text{Im}\{\hat{a}_2\} = \ldots = \text{Im}\{\hat{a}_{m-1}\} = 0$. For $s_0 = \sigma_0$, we show (28) and (29) by taking the contour integral of the function

$$[(\sigma_0 - s)^{-(k+1)} + (\sigma_0 + s)^{-(k+1)}] \log \hat{f}(s) \qquad (38)$$

along the close curve shown in Figure 9(c). This function is analytic in the RHP except $\sigma_0$; it is also analytic on the imaginary axis except possible zeros and poles of $\hat{f}(s)$ on the imaginary axis, which we denote as $j\omega_\ell$. Therefore, the following contour integral is zero for $k = 0, 1, 2, \ldots, m-1$:

$$\int_{\mathcal{C}_1 + \mathcal{C}_2 + \mathcal{C}_3 + \mathcal{C}_4} [(\sigma_0 - s)^{-(k+1)} + (\sigma_0 + s)^{-(k+1)}]$$

$$\times \log \hat{f}(s) ds = 0, \qquad (39)$$

where $\mathcal{C}_1$ is the line segment between $-jR$ and $jR$ excluding $[j(\omega_\ell - \varepsilon), j(\omega_\ell + \varepsilon)]$; $\mathcal{C}_2$ is the right semicircle with radius $R$ centered at the origin; $\mathcal{C}_3$ is the circle with radius $\varepsilon$ centered at the $\sigma_0$; and $\mathcal{C}_4$ includes the right semicircles with radius $\varepsilon$ centered at $j\omega_\ell$.

To show (28), we evaluate the integral paths in (39) for $k = 0$, and let $R \to \infty$, $\varepsilon \to 0$. The result is

$$\int_{-\infty}^{\infty} [(\sigma_0 - j\omega)^{-1} + (\sigma_0 + j\omega)^{-1}] \log |f(j\omega)| d\omega$$
$$= 2\pi \log |\hat{c}|.$$

We have $[(\sigma_0 - j\omega)^{-1} + (\sigma_0 + j\omega)^{-1}] \log |f(j\omega)|$ is an even function, and using (32) we get (28).

To show (29), we evaluate the integral paths in (39) for $k = 1, 2, \ldots, m-1$, and let $R \to \infty$, $\varepsilon \to 0$. The result is

$$\int_{-\infty}^{\infty} [(\sigma_0 - j\omega)^{-(k+1)} + (\sigma_0 + j\omega)^{-(k+1)}] \log |f(j\omega)| d\omega$$
$$= (-1)^k 2\pi \hat{a}_k.$$

We have $[(\sigma_0 - j\omega)^{-(k+1)} + (\sigma_0 + j\omega)^{-(k+1)}] \log |f(j\omega)|$ is an even function, and using (33) we get (29) for $k = 1, 2, \ldots, m-1$. This finishes the proof for $s_0 = \sigma_0$.

*4) $s_0 = \sigma_0 + j\omega_0$:* For $s_0 = \sigma_0 + j\omega_0$, we show (28) and (29) by taking the contour integral of the following functions

$$[(s_0 - s)^{-(k+1)} + (s_0 + s)^{-(k+1)} + (s_0^* - s)^{-(k+1)}$$
$$+ (s_0^* + s)^{-(k+1)}] \log \hat{f}(s) \tag{40a}$$

$$[(s_0 - s)^{-(k+1)} + (s_0 + s)^{-(k+1)} - (s_0^* - s)^{-(k+1)}$$
$$- (s_0^* + s)^{-(k+1)}] \log \hat{f}(s) \tag{40b}$$

along the close curve shown in Figure 9(d). These functions are analytic in the RHP except $s_0$ and $s_0^*$; they are also analytic on the imaginary axis except possible zeros and poles of $\hat{f}(s)$ on the imaginary axis, which we denote as $j\omega_\ell$. Therefore, the following contour integral is zero for $k = 0, 1, 2, \ldots, m - 1$:

$$\int_{\mathcal{C}_1 + \mathcal{C}_2 + \mathcal{C}_3 + \mathcal{C}_4} [(s_0 - s)^{-(k+1)} + (s_0 + s)^{-(k+1)}$$
$$+ (s_0^* - s)^{-(k+1)} + (s_0^* + s)^{-(k+1)}] \log \hat{f}(s)ds = 0, \tag{41a}$$

$$\int_{\mathcal{C}_1 + \mathcal{C}_2 + \mathcal{C}_3 + \mathcal{C}_4} [(s_0 - s)^{-(k+1)} + (s_0 + s)^{-(k+1)}$$
$$- (s_0^* - s)^{-(k+1)} - (s_0^* + s)^{-(k+1)}] \log \hat{f}(s)ds = 0, \tag{41b}$$

where $\mathcal{C}_1$ is the line segment between $-jR$ and $jR$ excluding $[j(\omega_\ell - \varepsilon), j(\omega_\ell + \varepsilon)]$; $\mathcal{C}_2$ is the right semicircle with radius $R$ centered at the origin; $\mathcal{C}_3$ includes the circles with radius $\varepsilon$ centered at the $s_0$ and $s_0^*$; and $\mathcal{C}_4$ includes the right semicircles with radius $\varepsilon$ centered at $j\omega_\ell$.

To show (28), we evaluate the integral paths in (41a) for $k = 0$, and let $R \to \infty$, $\varepsilon \to 0$. The result is

$$\int_{-\infty}^{\infty} \text{Re}[(s_0 - j\omega)^{-1} + (s_0 + j\omega)^{-1}] \log |f(j\omega)| d\omega$$
$$= 2\pi \log |\hat{c}|.$$

We have $\text{Re}[(s_0 - j\omega)^{-1} + (s_0 + j\omega)^{-1}] \log |f(j\omega)|$ is an even function, and using (32) we get (28).

To show (29), we first evaluate the integral paths in (41) for $k = 1, 2, \ldots, m - 1$, and let $R \to \infty$, $\varepsilon \to 0$. The result is

$$\int_{-\infty}^{\infty} [(s_0 - j\omega)^{-(k+1)} + (s_0 + j\omega)^{-(k+1)}] \log |f(j\omega)| d\omega$$
$$= (-1)^k 2\pi \hat{a}_k.$$

We have $[(s_0 - j\omega)^{-(k+1)} + (s_0 + j\omega)^{-(k+1)}] \log |f(j\omega)|$ is an even function, and using (33) we get (29). This finishes the proof for $s_0 = \sigma_0 + j\omega_0$. ∎

*5) Remark on lemma 3:* Because the formulas (28), (29) subtract or divide poles from zeros in equal quantity, when we apply Lemma 3 to $f(s) = \det A(s)$, (28), (29) still hold if the poles and zeros of the determinant are replaced by the poles and zeros of the matrix $A(s)$.

### C. Preliminary lemmas on S-matrices

*Lemma 4:* If $\text{Re}\{s_0\} > 0$ and the Darlington network given in Lemma 1 is used, then

$$\det S_{GM}(s_0) = \det S_{b22}(s_0) \neq 0, \tag{42}$$

and

$$\left( \sum_i (p_{GM,i} - s_0)^{-\ell} - \sum_i (z_{GM,i} - s_0)^{-\ell} \right)$$

$$= \left( \sum_i (p_{b22,i} - s_0)^{-\ell} - \sum_i (z_{b22,i} - s_0)^{-\ell} \right) \tag{43}$$

for $\ell = 1, \ldots, m - 1$.

If $\text{Re}\{s_0\} = 0$, then $m$ is even, (42) holds, and (43) holds for $\ell = 1, \ldots, m - 2$.

If $\text{Re}\{s_0\} = 0$ and $I - S_L(s_0)S_G(s_0)$ is non-singular then (43) holds for $\ell = m - 1$.

If $\text{Re}\{s_0\} = 0$ and $I - S_L(s_0)S_G(s_0)$ is singular then

$$(-1)^{\frac{m}{2}} \times$$
$$\left( \sum_i (p_{GM,i} - s_0)^{-(m-1)} - \sum_i (z_{GM,i} - s_0)^{-(m-1)} \right)$$
$$\leq (-1)^{\frac{m}{2}} \times$$
$$\left( \sum_i (p_{b22,i} - s_0)^{-(m-1)} - \sum_i (z_{b22,i} - s_0)^{-(m-1)} \right). \tag{44}$$

*Proof:* We separate our discussion into two possibilities: $\text{Re}\{s_0\} > 0$ and $\text{Re}\{s_0\} = 0$.

*1) $\text{Re}\{s_0\} > 0$:* Because the Darlington network in Lemma 1 is assumed, $S_{b22}(s)$ has no zeros at $s = s_0$, for otherwise $S_L(s)$ would have zeros at $s = -s_0$, and therefore have poles at $s = s_0$ in order to satisfy (18). This contradicts $S_L(s)$ being Hurwitzian. Hence, $\det S_{b22}(s_0) \neq 0$, and $S_L(s)$ has no zeros at $s = -s_0$.

Since $S_b(s)$ is lossless and therefore $S_b^T(-s)S_b(s) = I$, we get $S_{b12}(s) = -S_L^{-T}(-s)S_{b21}^T(-s)S_{b22}(s)$. We manipulate (15) to get

$$S_{GM}(s) = [I - S_{b21}(s)S_G(s)(I - S_L(s)S_G(s))^{-1}$$
$$\times S_L^{-T}(-s)S_{b21}^T(-s)]S_{b22}(s).$$

Taking determinant on both sides yields

$$\det S_{GM}(s) = \det S_{b22}(s) \det[I - S_{b21}(s)S_G(s)$$
$$\times (I - S_L(s)S_G(s))^{-1}S_L^{-T}(-s)S_{b21}^T(-s)]$$
$$= \det S_{b22}(s) \det[I - S_{b21}^T(-s)S_{b21}(s)S_G(s)$$
$$\times (I - S_L(s)S_G(s))^{-1}S_L^{-T}(-s)].$$

Since $S_L(s)$ and $S_G(s)$ are bounded, $S_L(s)S_G(s)$ is also bounded and $I - S_L(s_0)S_G(s_0)$ is non-singular for $\text{Re}\{s_0\} > 0$ [24, 7.22]. Hence, $S_{b21}^T(-s)S_{b21}(s) = I - S_L^T(-s)S_L(s) = O((s - s_0)^m)$, and $S_L(-s_0)$ and $I - S_L(s_0)S_G(s_0)$ are non-singular. We then have

$$\det S_{GM}(s) = \det S_{b22}(s)[1 + O((s - s_0)^m)].$$

Thus $\det S_{b22}(s_0) = \det S_{GM}(s_0) \neq 0$, and (42) holds for $\text{Re}\{s_0\} > 0$.

To show (43), we apply Lemma 2 to $\det S_{b22}(s)$ and $\det S_{GM}(s)$:

$$\log \det S_{GM}(s) = a_0 + a_1(s - s_0) + \ldots$$
$$+ a_{m-1}(s - s_0)^{m-1} + \ldots$$
$$\log \det S_{b22}(s) = b_0 + b_1(s - s_0) + \ldots$$
$$+ b_{m-1}(s - s_0)^{m-1} + \ldots, \tag{45}$$

where $a_\ell$ and $b_\ell$ have the form (27). Because $\det S_{GM}(s) = \det S_{b22}(s) + O((s - s_0)^m)$, $a_\ell = b_\ell$ for $\ell = 0, 1, \ldots, m - 1$. Writing $a_\ell$ and $b_\ell$ in the form of (27) yields (43).

*2) Re{s_0} = 0:* Let $s_0 = j\omega_0$. Since $S_b(s)$ is lossless and (18) is satisfied, $S_L^H(j\omega_0)S_L(j\omega_0) = S_{b22}^H(j\omega_0)S_{b22}(j\omega_0) = I$. Hence, $\det S_{b22}(j\omega_0) \neq 0$.

We begin by showing that $m$ is even. We substitute $s = j(\omega_0 \pm \varepsilon)$ into (18):

$$I - S_L^T(-j(\omega_0 \pm \varepsilon))S_L(j(\omega_0 \pm \varepsilon))$$
$$= I - S_L^H(j(\omega_0 \pm \varepsilon))S_L(j(\omega_0 \pm \varepsilon)) = A_m(\pm j\varepsilon)^m + \dots .$$

Since $S_L(s)$ is bounded, the $A_m(\pm j\varepsilon)^m$ is positive semidefinite. With $A_m \neq 0$, it is possible only when $m$ is even.

If $I - S_L(j\omega_0)S_G(j\omega_0)$ is non-singular, we follow a method similar to the Re{$s_0$} > 0 case to get $\det S_{GM}(s) = \det S_{b22}(s) + O((s-j\omega_0)^m)$. Thus (42) and (43) hold for $\ell = 1, 2, \dots, m-1$.

If $I - S_L(j\omega_0)S_G(j\omega_0)$ is singular, (44) can be proven in a manner similar to the proof of Lemma 5 in [9]. These steps are omitted. This finishes the proof of Lemma 4. ∎

### D. Proof of Theorem 2

We use the Darlington network in Lemma 1. Then Lemma 4 gives $\det S_{b22}(s_0) = \det S_{GM}(s_0) \neq 0$. Because $S_b(s)$ is para-unitary, (18) implies $\det(S_{b22}^T(-s)S_{b22}(s)) = \det(S_L^T(-s)S_L(s)) = 1 + O((s-s_0)^m)$. Hence, $\det S_{b22}(s)$ satisfies the conditions of Lemma 3.

From Lemma 4, $\det S_{GM}(s) = \det S_{b22}(s) + O((s-s_0)^m)$ for Re{$s_0$} > 0, and $\det S_{GM}(s) = \det S_{b22}(s) + O((s-s_0)^{m-1})$ for Re{$s_0$} = 0. When Re{$s_0$} = 0, we let $s_0 = j\omega_0$ and consider $s = j(\omega_0 \pm \varepsilon)$ for $\varepsilon > 0$:

$$\det[S_{GM}^T(-j(\omega_0 \pm \varepsilon))S_{GM}(j(\omega_0 \pm \varepsilon))]$$
$$= \det[S_{GM}^H(j(\omega_0 \pm \varepsilon))S_{GM}(j(\omega_0 \pm \varepsilon))]$$
$$= 1 + b_{m-1}(\pm j\varepsilon)^{m-1} + O(\varepsilon^m) \leq 1.$$

The inequality is because $S_{GM}(s)$ is bounded. Since $m-1$ is odd, $b_{m-1} = 0$. Hence, $\det(S_{GM}^T(-s)S_{GM}(s)) = 1 + O((s-s_0)^m)$, and $S_{GM}(s)$ satisfies the conditions of Lemma 3.

Unfortunately, Lemma 3 does not apply to $\det S_L(s)$ since there are possible zeros of $S_L(s)$ at $s = s_0$ when Re{$s_0$} > 0. From (18), if $S_L(s)$ has zeros at $s_0$, it also has zeros at $s_0^*$ and poles at $-s_0, -s_0^*$; the multiplicities of these poles or zeros are equal. We therefore construct a function $\widehat{\det S_L}(s)$ by removing the poles at $-s_0, -s_0^*$ and zeros at $s_0, s_0^*$ from $\det S_L(s)$, such that $|\widehat{\det S_L}(j\omega)| = |\det S_L(j\omega)|$. This function satisfies $\widehat{\det S_L}(-s)\widehat{\det S_L}(s) = 1 + O((s-s_0)^m)$ and $\widehat{\det S_L}(s_0) \neq 0$, thus the conditions of Lemma 3.

Since $S_b(s)$ is para-unitary, $|\det S_L(j\omega)| = |\det S_{b22}(j\omega)|$. It follows that

$$|\det S_L(j\omega)| = |\widehat{\det S_L}(j\omega)| = |\det S_{b22}(j\omega)|. \tag{46}$$

We first apply (28):

$$\int_0^\infty \text{Re}[(s_0 - j\omega)^{-1} + (s_0 + j\omega)^{-1}] \log |\det S_{GM}(j\omega)|d\omega$$
$$= \pi \log \left| \det S_{GM}(s_0) \cdot \frac{\prod_i(s_0 + z_{GM+,i})}{\prod_i(s_0 - z_{GM+,i})} \right| \tag{47a}$$

$$\int_0^\infty \text{Re}[(s_0 - j\omega)^{-1} + (s_0 + j\omega)^{-1}] \log |\det S_{b22}(j\omega)|d\omega$$

$$= \pi \log \left| \det S_{b22}(s_0) \cdot \frac{\prod_i(s_0 + z_{b22+,i})}{\prod_i(s_0 - z_{b22+,i})} \right| \tag{47b}$$

$$\int_0^\infty \text{Re}[(s_0 - j\omega)^{-1} + (s_0 + j\omega)^{-1}] \log |\widehat{\det S_L}(j\omega)|d\omega$$
$$= \pi \log \left| \widehat{\det S_L}(s_0) \cdot \frac{\prod_i(s_0 + z_{L+,i})}{\prod_i(s_0 - z_{L+,i})} \right|. \tag{47c}$$

Note $s_0$ and $s_0^*$ are excluded in $z_{L+,i}$ in (47c). From the definition of $\widehat{\det S_L}(s)$, we can rewrite (47c) as

$$\int_0^\infty \text{Re}[(s_0 - j\omega)^{-1} + (s_0 + j\omega)^{-1}] \log |\widehat{\det S_L}(j\omega)|d\omega$$
$$= \pi \log \left| \det S_L(s_0) \cdot \frac{\prod_i(s_0 + z_{L+,i})}{\prod_i(s_0 - z_{L+,i})} \right|, \tag{48}$$

with $s_0$ and $s_0^*$ included in $z_{L+,i}$. Because of (46), the integral in (47b) is the same if we replace $|\det S_{b22}(j\omega)|$ with $|\widehat{\det S_L}(j\omega)|$. Hence, the right-hand sides of (47b) and (48) are equal. We now apply (42) in Lemma 4 to the right-hand sides of (47a) and (47b), and (17) in Lemma 1 to the right-hand side of (47b). The result is (20).

We then apply (29):

$$\int_0^\infty [(s_0 - j\omega)^{-(k+1)} + (s_0 + j\omega)^{-(k+1)}]$$
$$\times \log |\det S_{GM}(j\omega)|d\omega$$
$$= \frac{(-1)^k \pi}{k} \Big[ \sum_i (p_{GM,i} - s_0)^{-k} - \sum_i (z_{GM,i} - s_0)^{-k}$$
$$+ \Big( \sum_i (z_{GM+,i} - s_0)^{-k} - \sum_i (-z_{GM+,i} - s_0)^{-k} \Big) \Big] \tag{49a}$$

$$\int_0^\infty [(s_0 - j\omega)^{-(k+1)} + (s_0 + j\omega)^{-(k+1)}]$$
$$\times \log |\det S_{b22}(j\omega)|d\omega$$
$$= \frac{(-1)^k \pi}{k} \Big[ \sum_i (p_{b22,i} - s_0)^{-k} - \sum_i (z_{b22,i} - s_0)^{-k}$$
$$+ \Big( \sum_i (z_{b22+,i} - s_0)^{-k} - \sum_i (-z_{b22+,i} - s_0)^{-k} \Big) \Big] \tag{49b}$$

$$\int_0^\infty [(s_0 - j\omega)^{-(k+1)} + (s_0 + j\omega)^{-(k+1)}]$$
$$\times \log |\widehat{\det S_L}(j\omega)|d\omega$$
$$= \frac{(-1)^k \pi}{k} \Big[ \sum_i (p_{L,i} - s_0)^{-k} - \sum_i (z_{L,i} - s_0)^{-k}$$
$$+ \Big( \sum_i (z_{L+,i} - s_0)^{-k} - \sum_i (-z_{L+,i} - s_0)^{-k} \Big) \Big], \tag{49c}$$

where $k = 1, 2, \dots, m-1$. Note $-s_0$ and $-s_0^*$ are excluded in $p_{L,i}$, and $s_0$ and $s_0^*$ are excluded in $z_{L,i}$ and $z_{L+,i}$ in (49c). Because of (46), the integral in (49b) is the same if we replace $|\det S_{b22}(j\omega)|$ with $|\widehat{\det S_L}(j\omega)|$. Hence, the right-hand sides of (49b) and (49c) are equal.

When $k \neq m-1$ or $I - S_L(s_0)S_G(s_0)$ is non-singular, we apply (43) in Lemma 4 to the right-hand sides of (49a) and (49b), and (17) in Lemma 1 to the right-hand side of (49b). The result is (21).

When $k = m - 1$ and $I - S_L(s_0)S_G(s_0)$ is singular, we apply (44) instead of (43) to the right-hand sides of (49a) and (49b). The result is (22). ∎

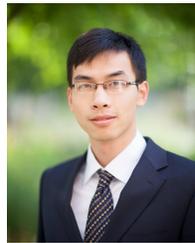

**Ding Nie** Ding Nie (M'16-) was born in Nanchang, Jiangxi, Peoples Republic of China. He received the B.S. degree in electronic engineering from Shanghai Jiao Tong University, Shanghai, in 2010. He received the M.S. degree and the Ph.D. degree in electrical engineering from the University of Notre Dame, Notre Dame, IN, in 2016. He then joined Apple Inc., Cupertino, CA, USA, in 2016. He won the 2016 Outstanding Young Author Award for the paper he co-authored in the IEEE Transactions on Circuits and Systems. His research interests include communications, radio-frequency circuits and antennas.

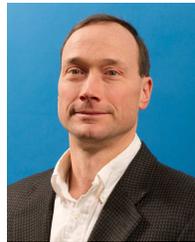

**Bertrand M. Hochwald** Bertrand M. Hochwald (F'08-) was born in New York, NY. He received his undergraduate education from Swarthmore College, Swarthmore, PA. He received the M.S. degree in electrical engineering from Duke University, Durham, NC, and the M.A. degree in statistics and the Ph.D. degree in electrical engineering from Yale University, New Haven, CT.

From 1986 to 1989, he worked for the Department of Defense, Fort Meade, MD. After completing graduate school he was a Research Associate and Visiting Assistant Professor at the Coordinated Science Laboratory, University of Illinois, Urbana-Champaign. In September 1996, he joined the Mathematics of Communications Research Department at Bell Laboratories, Lucent Technologies, Murray Hill, NJ where he was a Distinguished Member of the Technical Staff.

In 2005 he joined Beceem Communications, Santa Clara, CA, as their Chief Scientist, and in 2009 as Vice-President of Systems Engineering. He served concurrently as a Consulting Professor in Electrical Engineering at Stanford University, Palo Alto, CA. In 2011, he joined the faculty at the University of Notre Dame as Freimann Professor of Electrical Engineering.

He received several achievement awards while employed at the Department of Defense and the Prize Teaching Fellowship at Yale University. He has served as an Editor for several IEEE journals and given plenary and invited talks on various aspects of signal processing and communications. He has co-invented several well-known multiple-antenna techniques, including a differential method, linear dispersion codes, and multi-user vector perturbation methods. He has forty-five patents.

He received the 2006 Stephen O. Rice Prize for the best paper published in the IEEE Transactions on Communications. He co-authored a paper that won the 2016 Best Paper Award by a young author in the IEEE Transactions on Circuits and Systems. He is listed as a Thomson Reuters Most Influential Scientific Mind in 2014 and 2015. He is a Fellow of the IEEE.